\documentclass[twocolumn]{aastex631}
\usepackage{amsmath}
\usepackage{color,comment}
\usepackage{graphicx,subfigure}

\begin{document}

\title{Gravitational instability in protoplanetary disk with cooling: 2D global analysis}

\correspondingauthor{Xing Wei}
\email{xingwei@bnu.edu.cn}

\author[0000-0001-5567-0309]{Zehao Su}
\affiliation{IFAA, School of Physics and Astronomy, Beijing Normal University, Beijing 100875, China}

\author[0000-0002-3641-6732]{Xing Wei}
\affiliation{IFAA, School of Physics and Astronomy, Beijing Normal University, Beijing 100875, China}

\begin{abstract}
Self-gravity is important in protoplanetary disks for planet formation through gravitational instability (GI). We study the cooling effect on GI in a thin two-dimensional protoplanetary disk. By solving the linear perturbation equations in global geometry, we obtain all the normal modes. Faster cooling leads to faster growth rate of GI with lower azimuthal wavenumber $m$. According to the spatial structure of normal modes at different mass-averaged Toomre number $\overline{Q}$ and dimensionless cooling timescale $\beta$, we identify three modes: local, transitional and global. The transitional modes are located in the outer disk while the other two modes in the inner disk. At $\beta\approx 1$ (the resonance of dynamical timescale and thermal timescale) the growth rate changes sharply and the transitional modes dominate. The disk $\alpha$ due to GI is much higher in the transitional modes than in the other two. Our result implies that the transitional modes at $\overline{Q}\approx 1$ and $\beta\approx 1$ can plausibly interpret the substructures and planet/brown dwarf formation in the outer disk.
\end{abstract}

\section{Introduction} \label{Introduction}

Gravitational instabilities (GI) play a significant role in both the evolution of protoplanetary disks (PPDs) and the planet formation process. \cite[e.g.,][]{Cameron1978M&P....18....5C,Boss1997,Durisen2007prpl.conf..607D,Boley_2010,Paardekooper2012MNRAS.421.3286P,Xu2021}. They likely manifest during the early stages of protostellar disk evolution when the disk mass occupies more than 10\% protostellar mass \citep{AdamsLin1993prpl.conf..721A,Armitage_2020}. Disk mass estimates suggest that $50\%$ of Class 0 and $25\%$ of Class I disks exhibit GI instability \citep{Kratter2016ARA&A..54..271K}.

The observational evidence suggests that GI may play a substantial role in PPDs. Recent advancements in observational facilities and instruments, such as ALMA, VLT/SPHERE, VLT/CRIRES, and GPI, have enabled spatially resolved observations of PPDs. Spiral structures have been detected in the sub-millimeter thermal emission reflected by the mm-sized dusts around the disk midplane \cite[e.g.,][]{Perez2016Sci...353.1519P,Huang2018ApJ...869L..43H,Kurtovic2018ApJ...869L..44K}. These observations are biased tracers of gas density profiles due to finite aerodynamic coupling between gas and dust. Moreover, optical and near-infrared (NIR) high contrast scattered-light images, arising from starlight scattered by $\mu$m-sized dusts suspended in the disk surface, have also revealed spiral structures \cite[e.g.,][]{Garufi2013A&A...560A.105G,Wagner2015ApJ...813L...2W,Avenhaus2017AJ....154...33A,Benisty2017A&A...597A..42B,Uyama2018AJ....156...63U}. The origin of these spirals remains a debate, with GI offering a plausible explanation \citep{Perez2016Sci...353.1519P,Meru2017ApJ...839L..24M,Tomida2017ApJ...835L..11T,Hall2018MNRAS.477.1004H,Huang2018ApJ...869L..43H,Speedie2024Natur.633...58S}. Additionally, the presence of substellar companions at large radii (with semimajor axes $a > 10$ AU) implies the importance of GI \citep[e.g.][]{Kratter2010ApJ...710.1375K,Forgan2013MNRAS.432.3168F,Forgan2015MNRAS.447..836F}.

The study of GI has a long history dated back to James Jeans. Regarding disk GI, in recent years, \cite{Goodman1988} explored the fundamental modes in self-gravitating incompressible tori, demonstrating the existence of I mode (with corotation outside the tori) and J mode (with corotation at the density maximum). \cite{Adams1989} showed that one-armed density wave can be unstable when the disk mass is sufficiently high ($M_{d}/M_{*}=1$). \cite{Noh1991ApJ...383..372N} found nonaxisymmetric instability of high $m$'s when the disk is not so massive ($M_{d}=0.05\sim0.5 M_{\odot}$). \cite{Hadley2014Ap&SS.353..191H} studied the global nonaxisymmetric instabilities in a thick disk to produce a large parameter space of star-disk system. \cite{Lodato201510.1111/j.1365-2966.2005.08875.x} revealed that sufficiently massive disks can exhibit short-lived arms. The series of studies \citep{Laughlin1996,Dong2015,Hall2019ApJ...871..228H,Chen2021} demonstrated that GI-induced spirals can have more than two arms. In addition to classical GI, drag-mediated two-fluid GI \citep{Coradini1981A&A....98..173C,Longarini2023MNRAS.519.2017L} and secular GI \citep{Ward2000orem.book...75W,Youdin2005astro.ph..8659Y,Shariff2011ApJ...738...73S,Michikoshi2012ApJ...746...35M,Takeuchi2012ApJ...749...89T,Tominaga2020ApJ...900..182T,Pierens2021MNRAS.504.4522P} have been proposed for less massive disks.

The dynamical consequences of GI are strongly influenced by disk thermodynamics. When a disk becomes gravitationally unstable, GI-induced spirals spread across the disk, forming shocks that heat the disk to increase thermal pressure against GI. However, radiative cooling can help remove heat from the disk, making it more unstable. The final state is determined by the balance between heating and cooling. When the cooling timescale is shorter than the dynamical timescale, fragmentation may occur, leading to the formation of gravitationally bound objects \cite[e.g.,][]{Boss2001ApJ...563..367B,Gammie2001,Johnson2003ApJ...597..131J,Rice2005MNRAS.364L..56R,Baehr2017ApJ...848...40B}. When the cooling timescale is larger than the dynamical timescale, the disk may settle into a quasi-steady, marginally unstable state \cite[e.g.,][]{Goldreich1965,Lin1987MNRAS.225..607L,Gammie2001,Lodato2004MNRAS.351..630L,Boley2006ApJ...651..517B,Michael2012ApJ...746...98M,Steiman-Cameron2013ApJ...768..192S}, potentially sustaining long-lived substructures in the disk.

Motivated by the relationship of disk thermodynamics and GI, we perform a linear stability analysis to study the effects of cooling time for disk GI. In Section \ref{Methods} we derive the linear perturbation equations and give the numerical methods. In Section \ref{Results} we show some representative modes in different GI regimes. In Section \ref{Statistics} we focus on the relationship of dominant modes and disk cooling timescales. In Section \ref{observations} we discuss some implications and caveats, especially the disk $\alpha$. In Section \ref{Summary} and we summarize our results.

\section{Methods} \label{Methods}

\subsection{Linear perturbation equations} \label{Linear}

We consider an inviscid two-dimensional gas disk with self-gravity in cylindrical coordinates $(r,\phi)$. The equations of mass, momentum, internal energy and self-gravity are given
\begin{equation} \label{continuity}
\frac{\partial \Sigma}{\partial t}+\frac{1}{r} \frac{\partial(r \Sigma u_{r})}{\partial r} + 
\frac{1}{r} \frac{\partial(\Sigma u_{\phi})}{\partial \phi}=0, 
\end{equation}

\begin{equation} \label{momentum_r}
\frac{\partial u_{r}}{\partial t}+u_{r} \frac{\partial u_{r}}{\partial r}+\frac{u_{\phi}}{r} 
\frac{\partial u_{r}}{\partial \phi}-\frac{u_{\phi}^{2}}{r} 
= -\frac{G M_{*}}{r^{2}} - 
\frac{1}{\Sigma}\frac{\partial P}{\partial r}-\frac{\partial \Psi}{\partial r}, 
\end{equation}

\begin{equation} \label{momentum_psi}
    \frac{\partial u_{\phi}}{\partial t}+u_{r} \frac{\partial u_{\phi}}{\partial r}+\frac{u_{\phi}}{r} \frac{\partial u_{\phi}}{\partial \phi}+\frac{u_{\phi} u_{r}}{r}=-\frac{1}{r} \frac{\partial P}{\partial \phi}-\frac{1}{r} \frac{\partial \Psi}{\partial \phi}, 
\end{equation}

\begin{equation} \label{thermo}
    \frac{d e}{d t}+P \frac{d}{d t}\left(\frac{1}{\Sigma}\right)=\left(\frac{\partial e}{\partial t}\right)_{\mathrm{cool}},
\end{equation}

\begin{equation} \label{possion}
\nabla^{2} \Psi_{d}=4 \pi G \Sigma \delta(z),
\end{equation}
where $\Sigma, P, e, u_{r}, u_{\phi}$ represent, respectively, surface density, pressure, internal energy, radial and azimuthal velocities. Here $\Psi_{d}$ denotes self-gravity potential of the disk and $\delta(z)$ is Dirac function. The thermodynamic effects are implemented on the right-hand side of Eq. \eqref{thermo} using the same strategy as in \citet{Miranda2020}, with $e$ relaxed to a prescribed unperturbed value $e_{0}(r) = c_{\rm{ s,adi}}^2(r) / [\gamma(\gamma - 1)]$ ($\gamma$: adiabatic index, $c_{\rm{ s,adi}}$: adiabatic sound speed) on a cooling timescale $t_{\rm{c}}$
\begin{equation}
\left( \frac{\partial e}{\partial t}\right)_{\mathrm{cool}} = -\frac{e-e_{0}}{t_{\rm{c}}}.
\end{equation}
Here, $t_c$ is modeled with the standard $\beta$ cooling approximation \citep{Gammie2001}
\begin{equation}
t_{\mathrm{cool}} = \beta\Omega_{K}^{-1}\
\end{equation}
where $\Omega_{K}$ is the Keplerian angular velocity.

We describe the physical quantities, including $\Sigma, P, e, u_{r}, u_{\phi}, \Psi_{d}$, in a general form $X = X_{0} + \delta X$, where the subscript $X_{0}$ denotes equilibrium and $\delta X$ perturbation. In the following texts we drop the subscript `0' of the unperturbed variables for simplicity. The perturbations are assumed to be Fourier harmonics
\begin{equation} \label{Fourier}
\delta X(r, \phi, t)=\delta X(r) \exp \left[i\left(\omega_{m} t - m\phi\right)\right].
\end{equation}
Here $m$ represents the azimuthal wave number, and $\omega_{m}$ is the complex wave frequency that can be written as $\omega_{m}= m \omega_{p} - i\gamma_{m} $ where $\omega_{p}$ is the oscillation rate (or pattern speed) and $\gamma_{m}$ the growth rate. Substituting Eq. \eqref{Fourier} into \eqref{continuity}-\eqref{possion} while keeping only the first-order terms yields the linear perturbation equations
\begin{equation} \label{eqlin1}
-i \tilde{\omega} \delta \Sigma - \frac{1}{r} \frac{\partial}{\partial r}\left(r \Sigma 
\delta u_{r}\right)+\frac{i m \Sigma}{r} \delta u_{\phi}=0, 
\end{equation}
\begin{equation} \label{eqlin2}
-i \tilde{\omega} \delta u_{r} + 2 \Omega \delta u_{\phi} = \frac{1}{\Sigma} 
\frac{\partial}{\partial r} \delta P - \frac{1}{\Sigma^{2}} \frac{d P}{d r} \delta
 \Sigma + \frac{\partial}{\partial r} \delta \Psi_{m}, 
\end{equation}
\begin{equation} \label{eqlin3}
-i \tilde{\omega} \delta u_{\phi} - \frac{\kappa^{2}}{2 \Omega} \delta u_{r}=
-\frac{i m}{r}\left(\frac{\delta P}{\Sigma}+\delta \Psi_{m}\right), 
\end{equation}
\begin{equation} \label{eqlin4}
-\left(\frac{1}{t_{\mathrm{c}}}+i \tilde{\omega}\right) \delta P
+\left(\frac{1}{\gamma t_{\mathrm{c}}}+i \tilde{\omega}\right) c_{\mathrm{s},
 \mathrm{adi}}^{2} \delta \Sigma = \frac{\Sigma c_{\mathrm{s}, \mathrm{adi}}^{2}}{L_{S}} \delta u_{r}
\end{equation} 
\begin{equation} \label{eqlin5}
\delta\Psi_{m}(r)=-\int_{r_0}^{r_d} 
d \rho \int_{0}^{2 \pi} \frac{G \cos (m \phi) \delta\Sigma(\rho) \rho}{\sqrt{\rho^{2}+r^{2}-2 \rho r\cos \phi} } d \phi .
\end{equation}
Here $\tilde{\omega} = \omega_{m} - m \Omega $ is the Doppler-shifted frequency of the perturbation where $\Omega(r)$ is the angular velocity of the unperturbed state. $L_{S}$ is the length scale of entropy variation, defined as $1/L_{S} = (1/\gamma) (d S/d r)$ where $S$ is the gas entropy and the adiabatic index $\gamma$ is given to be $7/5$. The perturbed self-gravitational potential is rewritten in its integral form \citep{Shu1992} and the range $[r_0,r_d]$ will be discussed in the next subsection. The angular velocity of unperturbed state can be derived from the radial equilibrium
\begin{equation}
\Omega=\sqrt{\frac{1}{r}\left(\frac{G M_{*}}{r^{2}}+\frac{1}{\Sigma}\frac{d P}{d r}+\frac{d \Psi}{d r}\right)},
\end{equation}
with the epicyclic frequency
\begin{equation}
\kappa^{2}=\frac{1}{r^{3}} \frac{d}{d r}\left[\left(r^{2} \Omega\right)^{2}\right].
\end{equation}

\subsection{Disk profiles}

As is known, self-gravity becomes crucial and GI can be triggered when $M_{d}/M_{*}$ exceeds 0.1 \citep{Dong2015,Kratter2016ARA&A..54..271K}. To specifically investigate GI in the early stages of disks, we deliberately choose a relatively substantial disk-star mass ratio of $M_{d}/M_{*}=0.4$. Usually the Toomre number $Q=\kappa c_{s}/\pi G \Sigma$ is used as the criterion of GI, i.e., a disk is susceptible to gravitational instability when $Q<1$ \citep{Toomre1964}. In our global analysis we employ the mass-averaged Toomre number $\overline{Q} = \int Q \Sigma dS/\int \Sigma dS$ rather than its local definition. We ensure $\overline{Q}\lesssim 2$ to trigger the nonaxisymmetric GI \citep{Chen2021}.

The exact profile of PPDs in their early stages remains elusive, despite advancements in observational facilities like ALMA and VLT. We acknowledge the challenge in constraining these profiles and, for simplicity, adopt straightforward and interpretable profiles for both surface density ($\Sigma$) and temperature ($T$). 

\begin{figure*}
    \centering
    \includegraphics[width=0.8\textwidth]{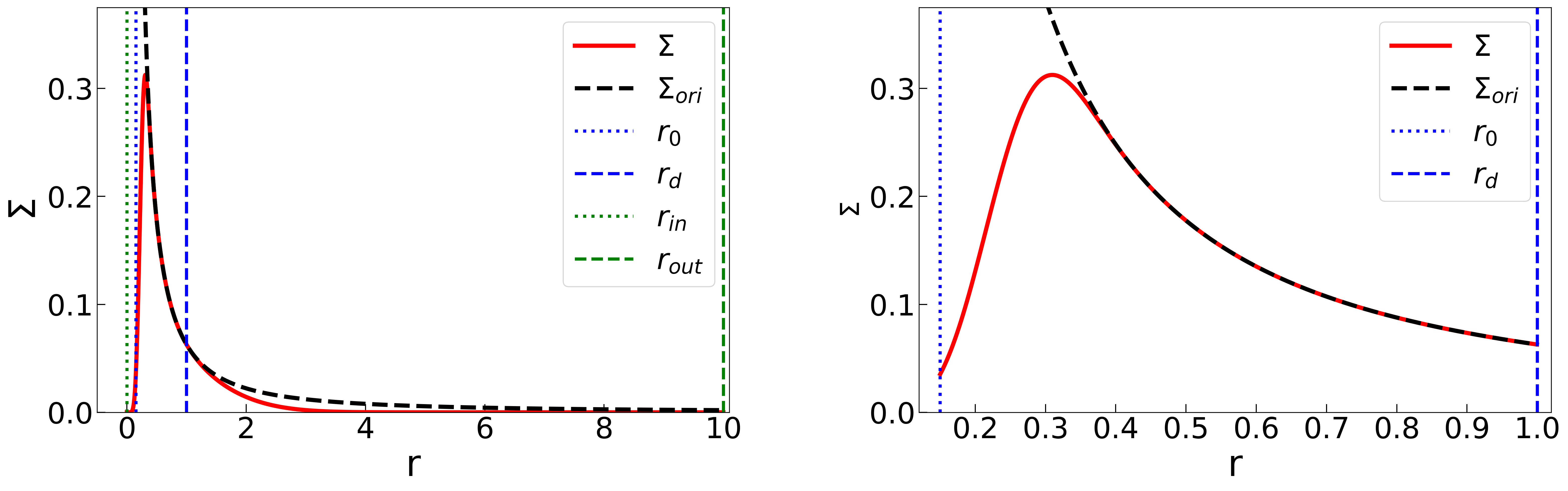}
    \caption{Surface density profiles within $[r_{\rm{in}}, r_{\rm{out}}]$ (left panel) and $[r_{0}, r_{d}]$ (right panel). The red solid line represents the density profile used in our calculation, while the black dashed line shows the profile $\Sigma_{\rm{ori}} \propto r^{-1.5}$. The green dotted and dashed lines indicate the positions of $r_{\rm{in}}$ and $r_{\rm{out}}$, respectively. The blue dotted and dashed lines denote the locations of $r_{0}$ and $r_{d}$, respectively.}
    \label{surface_density}
\end{figure*}

We assume the surface density profile
\begin{equation}
\Sigma(r) = C_{\Sigma} f_{\rm{tap}}(r) r^{-p} f_{\rm{trunc}}(r)
\end{equation}
where the constant $C_{\Sigma}$ is determined with disk mass. To suppress the abrupt change of the self-gravitational potential at the inner and outer boundaries of the disk which will induce the edge modes, we apply tapering ($f_{tap}$) and truncation ($f_{\rm{trunc}}$) functions. The tapering function $f_{\rm{tap}}$ is a polynomial with its order higher than 10 defined to be $\in[0,1]$ at the origin and the inner boundary $r_{0}$ \citep{Backus2016MNRAS.463.2480B}. The truncation function is defined for $r>r_{d}$ 
\begin{equation}
f_{trunc}(r) = e^{ -(r-r_{d})^{2}/L^{2} }
\end{equation}
where $r_{d}$ is the outer boundary and $L=0.15 r_{d}$. Both the tapering and truncation functions are chosen to ensure that the surface density decreases rapidly when $r$ extends beyond the calculation domain $r \in [r_{0},r_{d}]$. They also smoothly approach $0$ near $r={0,r_{out}}$ (where $r_{out}$ is the outer boundary used to calculate the equilibrium profile), while maintaining $\Sigma$ and $d\Sigma/dr$ unchanged at $r={r_{0},r_{d}}$. The power-law index $p$ for the surface density is set to $1.5$. Figure \ref{surface_density} illustrates the density profile of our equilibrium state.

We assume a power-law temperature profile
\begin{equation}
T(r) = C_{T} r^{-q}
\end{equation}
where the normalization factor $C_{T}$ is determined by $\overline{Q}$. We choose $q=0.6$, which is an intermediate value between $q=3/7$ given by the balance of disk cooling and stellar irradiation and $q=3/4$ given by the balance of disk cooling and accretion heating \citep{Kratter2016ARA&A..54..271K}.

\subsection{Boundary conditions}
The perturbation equations consist of two first-order differential equations so that we need two boundary conditions. For the inner boundary, we use the reflecting boundary condition, i.e., the perturbed radial velocity vanishes when approaching the inner boundary of disk
\begin{equation} \label{IBC}
    \delta u_{r} = 0  \quad {\rm at} \quad r=r_0.
\end{equation} 
For the outer boundary, we use the confining boundary condition, i.e., the Lagrangian perturbation of pressure vanishes at the outer boundary of disk
\begin{equation} \label{OBC}
    im \Omega \delta P - \frac{dP}{d r} \delta u_{r} = i \omega_{m} \delta P  \quad {\rm at} \quad r=r_d.
\end{equation}
We refer these boundary condition as 'RC' boundary. We also test the other boundary conditions `RR', `CC' and `CR', and find that the boundary conditions cannot substantically change eigenfrequencies and eigenfunctions in the disk interior.

\subsection{Numerical method} \label{Numerical}
We employ two different ranges in our analysis. The equilibrium is within $[r_{in},r_{out}]$ to avoid self-gravitational divergence at the boundaries and obtain a consistent rotation profile, whereas the perturbation in $[r_{0},r_{d}]$, with $r_{in}<r_{0}<r_{d}<r_{\rm{out}}$. In other words, the disk range is $[r_{0},r_{d}]$, and the regions between $[r_{in},r_{0})$ and $(r_{d},r_{\rm{out}}]$ serve solely as ghost ranges to obtain the equilibrium profile. The system is normalized with $G=M_{*}=\left.\Omega_{k}\right\vert_{r=r_{d}}=1$. We set $r_{in}=0$, $r_{0}=0.15$, $r_{d}=1$, and $r_{out}=10$ in our calculation.

We translate the integro-differential equations \eqref{eqlin1}-\eqref{eqlin5} to an eigenvalue problem, which has been widely used for searching normal modes in PPDs \citep{Adams1989,Noh1991ApJ...383..372N}. We give a brief introduction here and the details can be found in Appendix \ref{App1}. The eigenvalue problem is
\begin{equation} \label{eigen}
    (\boldsymbol{M_{1}}+\boldsymbol{M_{2}})\boldsymbol{X}=\boldsymbol{i\omega_{m}}\boldsymbol{X}
\end{equation}
where $\boldsymbol{X}$ represents the eigenvector and $\boldsymbol{i\omega_{m}}$ the eigenvalue. $\boldsymbol{M_{1}}$ and $\boldsymbol{M_{2}}$ are, respectively, the coefficient matrix for \eqref{eqlin1}-\eqref{eqlin5} without self-gravity terms and the coefficient matrix with only $\delta\Psi_{m}$ terms. We focus on the low $m\le 5$, as GI is favored by low azimuthal wavenumbers and the multiple-arm spirals ($>2$) are not often observed \citep{Bae2023ASPC..534..423B}. Only the modes with the highest growth rate are selected, as they dominate over the other modes in the linear regime.

We use the logarithm-spaced grids within $[r_{0},r_{d}]$ and take the resolution 800. To test the numerical convergence we double the resolution and the relative error of eigenfrequency is less than 2\%.

\section{Three regimes}  \label{Results}

\begin{figure*}
    \centering
    \includegraphics[width=0.75\textwidth]{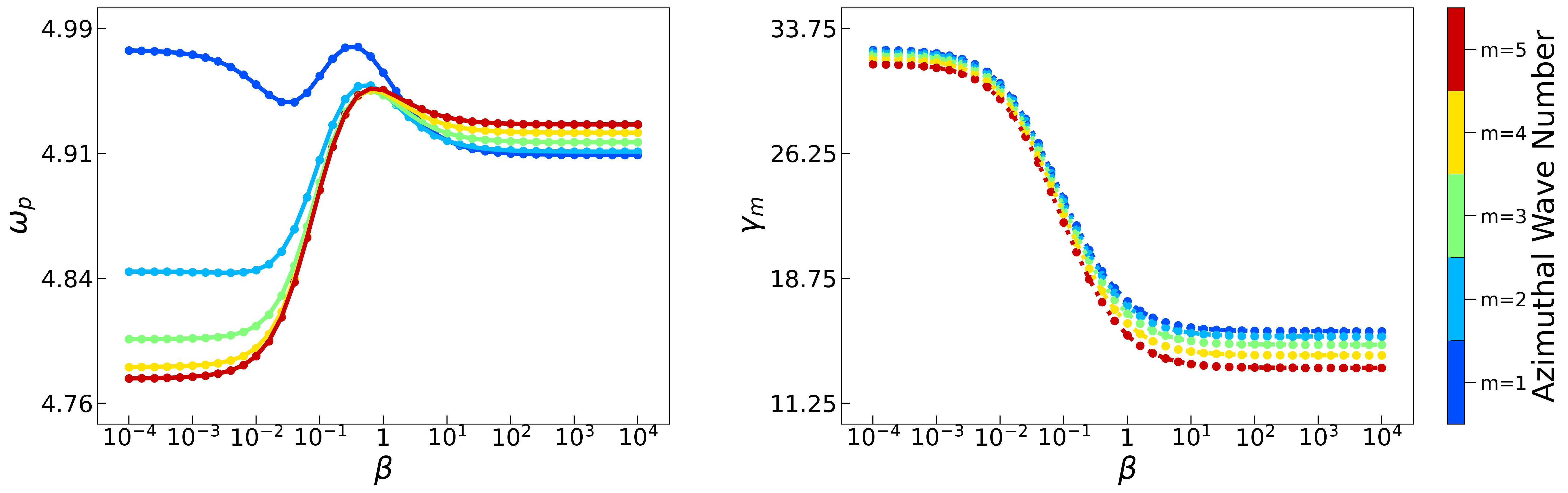}
    \caption{Oscillation frequency (left panel) and growth rate (right panel) versus cooling timescale $\beta$ for different azimuthal wave number $m$. Local regime at $\overline{Q}=0.53$. All wave frequencies are normalized by $\left.\Omega_{k}\right\vert_{r=r_{d}}$.}
    \label{low rate}
\end{figure*}

\begin{figure*}
    \centering
    \includegraphics[width=0.75\textwidth]{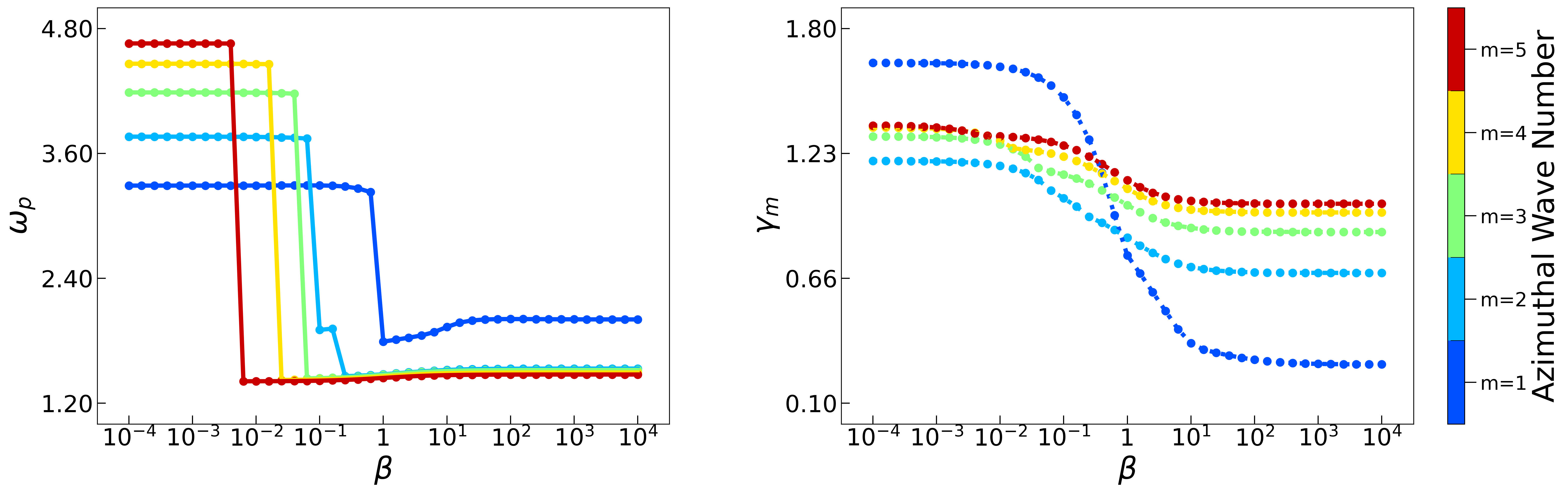}
    \caption{As in Fig.\ref{low rate} but transitional regime at $\overline Q=1.15$.}
    \label{transitional rate}
\end{figure*}

\begin{figure*}
    \centering
    \includegraphics[width=0.75\textwidth]{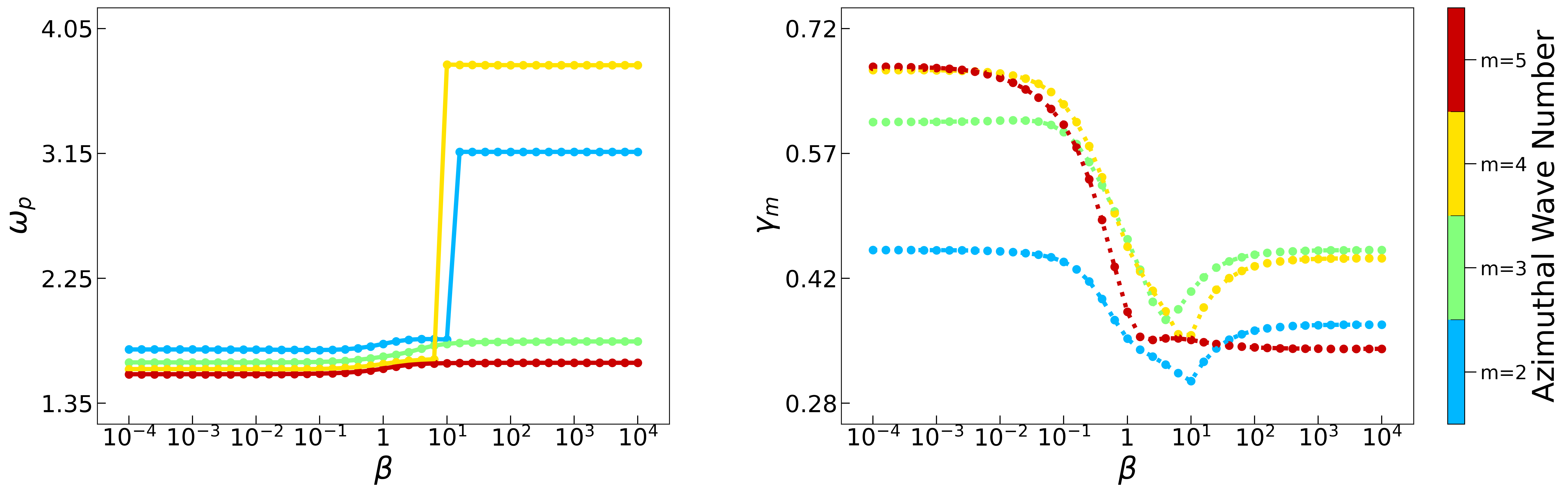}
    \caption{As in Fig.\ref{low rate} but global regime at $\overline Q=1.65$.}
    \label{high rate}
\end{figure*}

In this section we test with the mass-averaged Toomre numbers $\overline{Q}=0.53, 1.15, 1.65$ across the entire range of $\beta$ and show some representative results of GI modes. We focus on $m=2$ modes because the two-armed spirals are commonly observed in both millimeter continuum \citep{Andrews2018,Huang2018ApJ...869L..43H,Kurtovic2018ApJ...869L..44K} and near-infrared scattered-light observations \citep{Garufi2013A&A...560A.105G,Benisty2015A&A...578L...6B,Benisty2017A&A...597A..42B,Canovas2018A&A...610A..13C,Uyama2018AJ....156...63U}. To characterize the modes, we compare the morphology of eigenfunctions, the oscillation rate (or pattern speed) $\omega_{p}$, the growth rate $\gamma_{m}$, and the maximum location of $|\delta\Sigma|$ ($r_{|\delta\Sigma|_{\rm{max}}}$) under different $\overline{Q}$ and $\beta$. Additionally, we provide the locations of the inner and outer Lindblad resonances within which the spiral density wave propagates.

\subsection{Local regime} \label{Local}
We study the low Toomre number regime with $\overline{Q}=0.53$. The resulting eigenfunctions and eigenvalues are illustrated in Figs.\ref{low rate} and \ref{m=2 low}. In Figure \ref{m=2 low}, the left panels display the density perturbation as a function of radius, while the right panel shows the distribution of density perturbation on the disk plane. In the left panel, both real and imaginary parts of density perturbation are shown, and the phase angle between the two parts determines the spiral pattern depicted by the right panel.

--{\it Oscillation Rate}. In this regime, as shown in the left panel of Fig.\ref{low rate}, all modes have a pattern speed $\omega_{p}\in[4,5]$, indicating that their corotation radius (CR: vertical red line) are very close to the location of $\Sigma_{\rm{max}}$ where $\Omega=5.8$ ($r_{\Sigma_{\rm max}}=0.31$).

--{\it Growth Rate}. A notable feature of the growth rate $\gamma_{m}$ in this regime is that the dominant modes always own the smallest $m$, regardless of cooling timescale, and they exhibit extremely large growth rates ($>10$) compared to the other regimes discussed below. Another significant feature, as shown in the right panel of Fig.\ref{low rate}, is that $\gamma_{m}$ decreases as $\beta$ increases. This is because slow cooling (large $\beta$) leads to strong thermal pressure to oppose gravitational instability \citep{Jeans1902}. Our result is consistent with \cite{Gammie2001} that fast cooling triggers disk fragmentation. More interestingly, the growth rate changes sharply near $\beta\sim 1$ when orbital dynamical timescale is comparable to thermodynamic timescale, probably because of the resonance of orbital dynamics and thermodynamics.

--{\it Mode Morphology}. As shown in the left panels of Fig.\ref{m=2 low}, the real and imaginary parts of density perturbation exhibit the similar distribution, and the peaks are located near the maximum of density perturbation $r_{|\delta\Sigma|_{\rm{max}}}=0.36$ (vertical yellow line) as well as the corotation radius (vertical red line), and not far away from the maximum of equilibrium density $r_{\Sigma_{\rm max}}=0.31$ (vertical purple line). The right panels of Fig.\ref{m=2 low} clearly show the concentration behavior of these local modes. Moreover, the strongest density perturbation has the radial wave number $k \gg 1/r$ given by the WKBJ dispersion relation \citep{Binney2008}
\begin{equation}
    |k| = \frac{\pi G\Sigma}{c_{s}^{2}}\pm\frac{\pi G\Sigma}{c_{s}^{2}}\left\{ 1-Q^{2}\left[1-(\omega-m\Omega)^{2}/\kappa^{2}\right]\right\}^{1/2}.
\end{equation}

\subsection{Transitional regime} 
\label{transitional}

$Q\sim1$ is a threshold, below which the axisymmetric instability is triggered and slightly above which the nonaxisymmetric instability is triggered \citep{Kratter2016ARA&A..54..271K}. We take $\overline{Q}=1.15$ to investigate the GI behavior near this threshold, and the results are shown in Figs.\ref{transitional rate} and \ref{m=2 interm}. 

--{\it Oscillation Rate}. When $\beta$ gradually increases, the modes transit from the local regime to the transitional regime. A noticeable change occurs at $\beta \sim 1$(see the left panel of Fig.\ref{transitional rate}). The GI modes of all $m$'s enter the transitional regime when $\beta\gtrsim 1$. The typical $\omega_{p}$ in the transitional regime is around 1.4, which corresponds to the corotation radius (CR) in the outer disk.

\begin{figure}
    \centering
    \includegraphics[width=0.8\columnwidth]{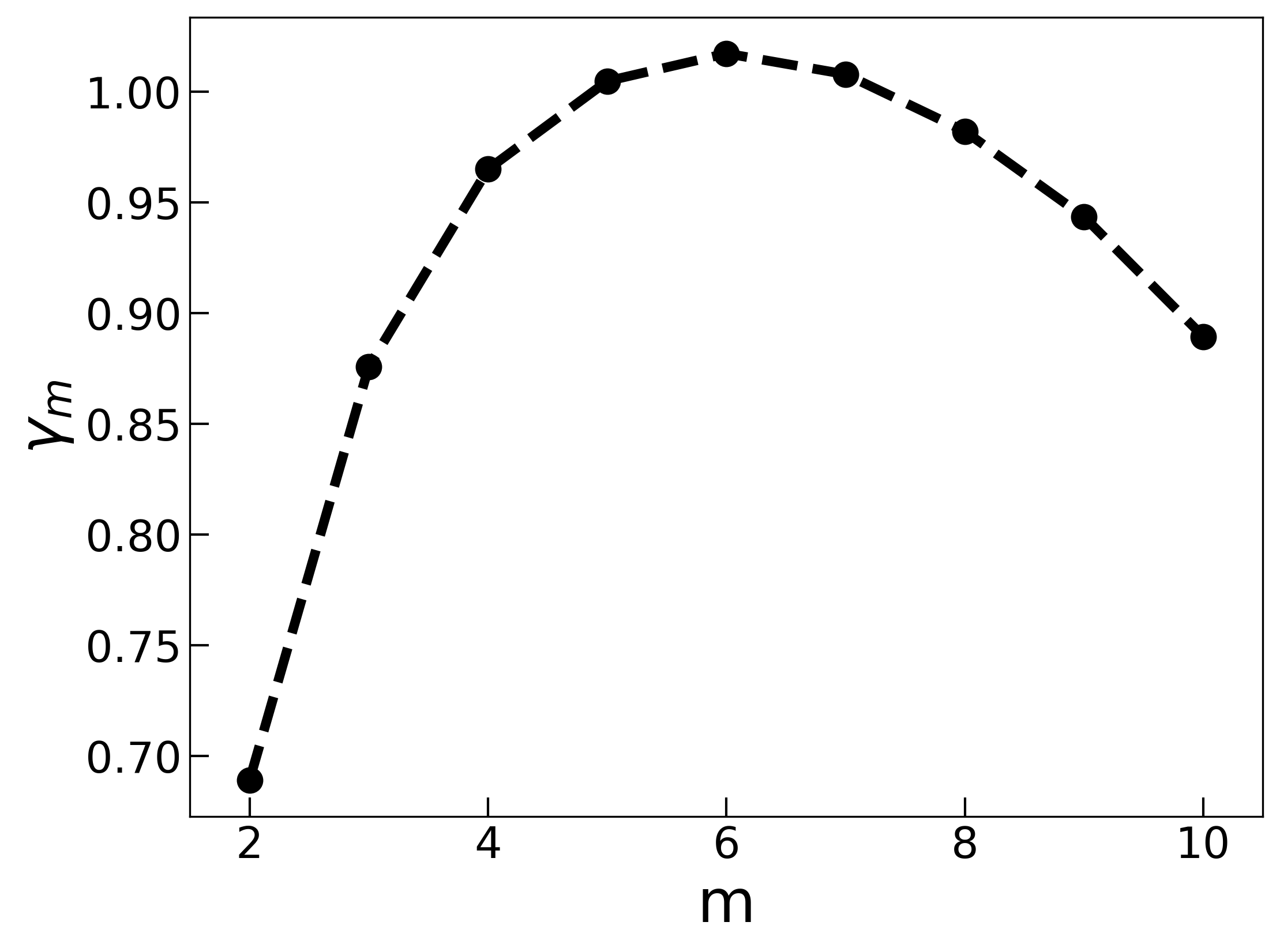}
    \caption{Growth rate $\gamma_{m}$ for $m=2-10$ modes at $\overline{Q}=1.15$ and $\beta=10^{4}$. The dominant mode is at $m=6$.}
    \label{transitional prove}
\end{figure}

--{\it Growth Rate}. Similar to the local modes, the transitional modes with faster cooling are more unstable, albeit the growth rates are one or two orders lower than the local modes because of higher $\overline Q$. Unlike the local modes in which the lowest $m=1$ always dominates, in the transitional modes the higher $m$ dominates when $\beta>1$. We also test even higher $m$ at $\beta=10^{4}$, shown in Fig.\ref{transitional prove}. The growth rate peaks at $m=6$ and then decreases as $m$ increases. This indicates that the modes are indeed caused by gravitational instability which prefers long wave.

--{\it Mode Morphology}. Compared to the local modes, the maximum density perturbations are also located near $r_{|\delta\Sigma|_{\rm{max}}}$ and the CR in the outer disk but far away from $r_{\Sigma_{\rm max}}$ as shown in the left panels of Fig.\ref{m=2 interm}, which is different from the local modes, and the density perturbation is much more extended in the radial direction as shown in the right panels of Fig.\ref{m=2 interm}, which is also different from the high concentration of the local modes.

\subsection{Global regime} \label{Global}

We study the high Toomre number regime $\overline{Q}=1.65$ which reaches the boundary of GI. Only the $m=2,3,4,5$ modes are calculated and the $m=1$ mode is stable. The results are shown in Figs.\ref{high rate} and \ref{m=2 high}.

--{\it Oscillation Rate}. The odd $m$'s lead to $\omega_{p}\sim 1.5$ corresponding to CR $\sim$ 0.76 in the outer disk. While the even $m$'s lead to $\omega_{p}\sim 1.5$ when $\beta<10$ but $\omega_{p}\sim 3$ when $\beta>10$ corresponding to CR $\sim$ 0.48 in the inner disk.

--{\it Growth Rate}. Similar to local and transitional modes, the growth rate $\gamma_{m}$ of global modes changes at $\beta\sim 1$ but it is lower than 1 because of high $\overline Q$. The dominant modes nearly follow $m\sim M_{*}/M_{d}$, consistent with the numerical simulations \cite{Dong2015}.

\begin{figure*}
    \centering
    \includegraphics[width=0.75\textwidth]{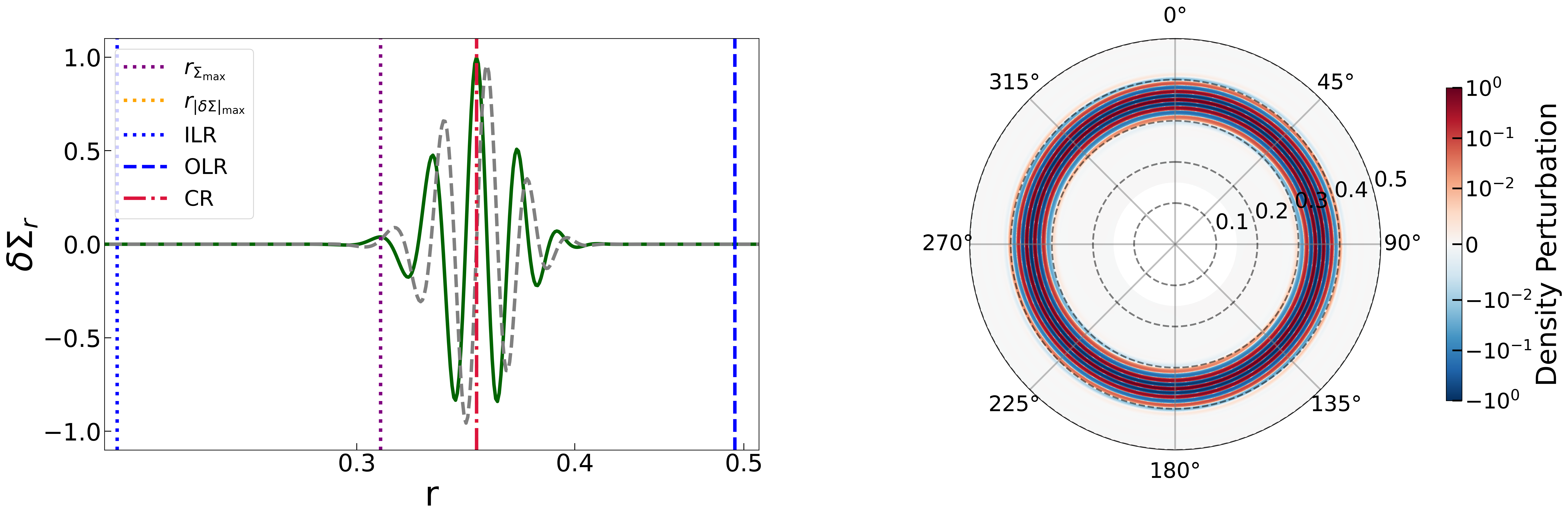}
    \caption{Morphology of the $m=2$ mode. Local regime at $\overline{Q}=0.53$ and $\beta=1$. The left panel shows the real part of density perturbation by green solid line, the imaginary part by gray dashed line, $\Sigma_{\rm{max}}$ by purple dashed line, and $r_{|\delta\Sigma|_{\rm{max}}}$ by yellow dashed line. The inner Lindblad resonance, outer Lindblad resonance and corotation resonance are marked by, respectively, blue dotted, blue dashed and red dashed lines. The right panel shows the ($r,\phi$) distribution of density perturbation.}
    \label{m=2 low}
\end{figure*}

\begin{figure*}
    \centering
    \includegraphics[width=0.75\textwidth]{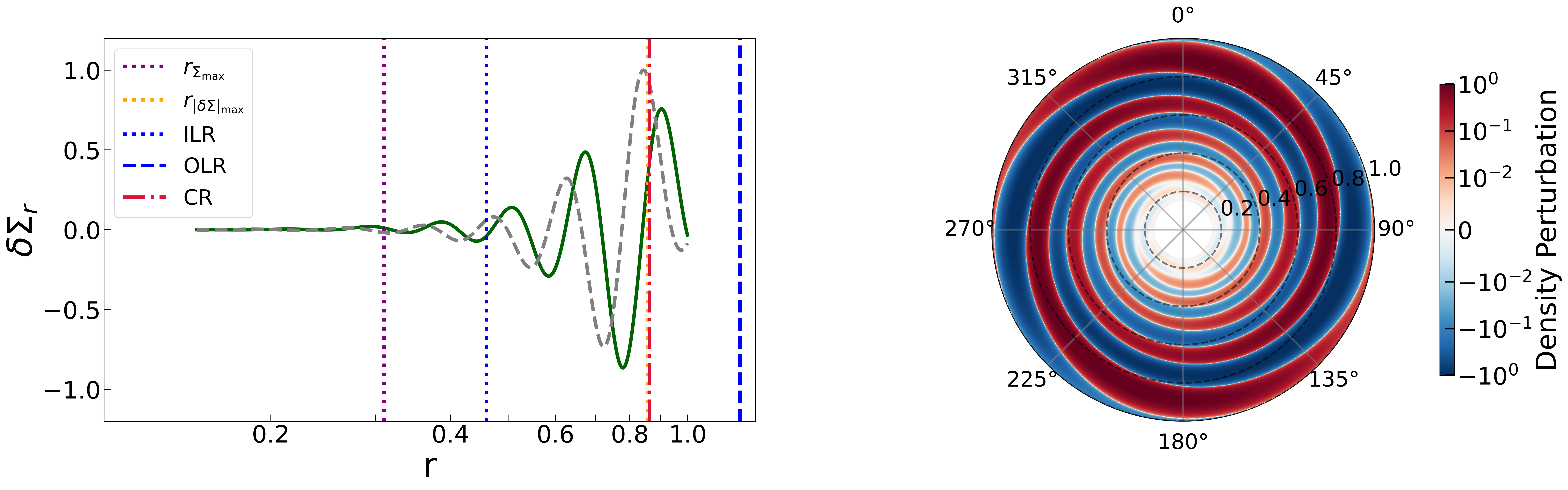}
    \caption{As in Fig.\ref{m=2 low} but transitional regime at $\overline Q=1.15$ and $\beta=1$.}
    \label{m=2 interm}
\end{figure*}

\begin{figure*}
    \centering
    \includegraphics[width=0.75\textwidth]{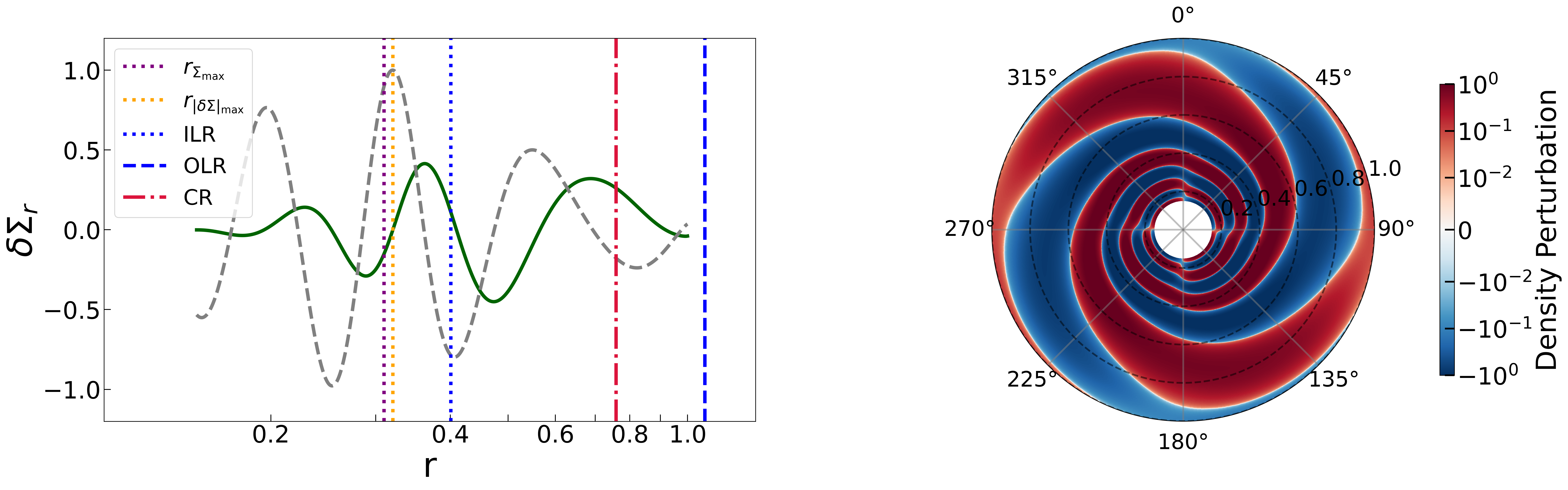}
    \caption{As in Fig.\ref{m=2 low} but global regime at $\overline Q=1.65$ and $\beta=10^{-2}$.}
    \label{m=2 high}
\end{figure*}

--{\it Mode Morphology}. As shown in Fig.\ref{m=2 high}, a notable feature in this regime is the global pattern, i.e., the perturbations are distributed across the entire disk. This results in fewer peaks compared to the other two regimes (see left panel), and the spirals now have significantly larger pitch angles (wider phase difference between real and imaginary parts, see left panel), making them more loosely winded (see right panel). Another significant difference from the other two regimes is that the location of maximum of density perturbation $r_{|\delta\Sigma|_{\rm{max}}}$ departs away from the CR location (red and yellow lines, see left panel), whereas in the other two regimes the two locations are close (see left panels of Figs.\ref{m=2 low} and \ref{m=2 interm}).

\section{Parameter scan}  \label{Statistics}

We have known that the growth rate and the mode transition depend not only on $\overline{Q}$ but also on $\beta$, and therefore, we scan the parameter space ($\beta,\overline Q$) for $\beta$ from $10^{-4}$ to $10^{4}$ and $\overline{Q}$ from $0.5$ to $1.7$ with fixed disk-star mass ratio $M_{d}/M_{*}=0.4$. We focus on the $m=2,3,4$ modes since $m=1$ mode may be stable and higher modes ($m\geqslant5$) may not be GI but shear instability (recall that GI prefers long wave).

\begin{figure*}
    \centering
    \includegraphics[width=0.95\textwidth]{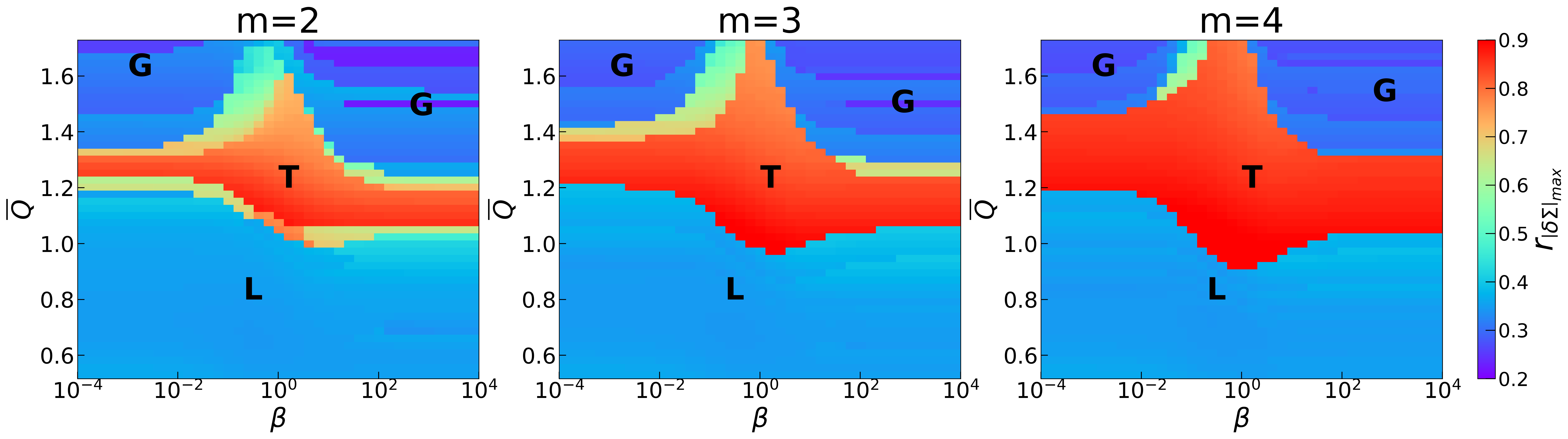}
    \caption{Parameter scan for $m=2,3,4$ with $\overline{Q}\in[0.5,1.7]$ and $\beta\in[10^{-4},10^{4}]$. The contour shows the location of $r_{|\delta\Sigma|{max}}$. The local regime marked by `L' dominates with low $\overline{Q}$, the transition regime marked by `T' occupies a large area for $\overline{Q}\gtrsim0.9$, and the global regime is observed with high $\overline{Q}$ when $\beta$ is away from unity.}
    \label{contrast regime}
\end{figure*}

We choose $r_{|\delta\Sigma|_{\rm{max}}}$ to distinguish the transitional regime ($r_{|\delta\Sigma|_{\rm{max}}} > 0.5$) from the local and global regimes ($r_{|\delta\Sigma|_{\rm{max}}} \lesssim 0.5$). Fig.\ref{contrast regime} shows the contour of $r_{|\delta\Sigma|_{\rm{max}}}$ as a function of $\beta$ and $\overline Q$. The bold characters ``$L,T,G$'' denote respectively the local, transitional and global regimes. For $\overline{Q}\lesssim 0.9$, the dominant modes in the disk are local regardless of disk thermodynamics $\beta$. As $\overline{Q}$ increases, at $\beta\sim1$ GI enters the transitional regime, and moreover, slower cooling ($\beta>1$) requires lower $\overline Q$ to enter the transitional regime than faster cooling ($\beta<1$). When $\overline{Q}$ continues to increase, GI with both large and small $\beta$ will eventually leave the transitional regime to enter the global regime. However, GI with $\beta\sim1$ tends to retain in the transitional regime, and higher mode owns a wider transitional regime. Similar to growth rate, this behavior at $\beta\sim1$ may also result from the resonance of orbital dynamics and thermodynamics.

\section{Implication for observations} \label{observations}

Gravitational instability significantly influences the evolution of PPDs and the planet formation. Firstly, ALMA has detected various substructures, e.g., spirals, rings, and gaps \citep{Andrews2018}, and GI is a potential mechanism for these large-scale substructures \citep{Lin1964,Binney2008,Bae2023ASPC..534..423B,Meru2017ApJ...839L..24M,Speedie2024Natur.633...58S}. Secondly, although core accretion is considered to be the classical model to understand the giant planet formation even at large distance $\gtrsim$ 10 AU \citep[e.g.,][]{Nielsen2019AJ....158...13N,Vigan2021A&A...651A..72V,Currie2023ASPC..534..799C}, GI is an alternative for the high-mass and long-period planet formation, especially to understand the metallicity non-correlation of brown dwarfs, debris disks, and their host stars \cite[e.g.,][]{Nayakshin_2017}. Thirdly, besides turbulent viscosity, GI can be responsible for the angular momentum transport \citep{Rice201710.1093/mnras/stx3255}.

The disk turbulent viscosity is related to accretion rate and angular momentum transport. Usually it is described by the disk $\alpha$ \citep{Shakura1973A&A....24..337S}
\begin{equation}
\nu = \alpha H c_{s}
\end{equation}
where $\nu$ is the turbulent viscosity, $\alpha$ is a dimensionless coefficient, $H$ is the disk scale height and $c_{s}$ is the local sound speed. The disk $\alpha$ is essentially the ratio of turbulent velocity of largest eddies to sound speed. The typical accretion rate $\sim 10^{-8}$ to $10^{-7} M_{\odot}$ in protoplanetary disks \cite[e.g.,][]{Armitage_2020} corresponds to $\alpha$ $\sim 10^{-3}$ to $10^{-2}$. Turbulent viscosity arises from Reynolds stress which is given by
\begin{equation}
T_{r\phi}^{\rm Rey}(r)=\frac{1}{2\pi}\int\Sigma\delta u_r\delta u_\phi d\phi.
\end{equation}
In addition to Reynolds stress, GI can also play a role in transporting angular momentum. Gravitational stress is given by \cite[e.g.,][]{Gammie2001,Lodato2004MNRAS.351..630L,Boley2006ApJ...651..517B,Michael2012ApJ...746...98M,Steiman-Cameron2023ApJ...958..139S}
\begin{equation}
T^{\rm{grav}}_{r\phi}(r) = -\frac{1}{2\pi r^{2}}\int \delta\Sigma \frac{\partial\delta\Psi}{\partial\phi} dS
\end{equation}
where $S$ is disk surface area. By Reynolds and gravitational stresses we can calculate the effective $\alpha$
\begin{equation}\label{eq:alpha}
\alpha = \left|\frac{T_{r\phi}}{\Sigma c_{s}^{2}}\left(\frac{d\ln\Omega}{d\ln r} \right)^{-1}\right|
\end{equation}
where the stress $T_{r\phi}$ can be either $T^{\rm Rey}_{r\phi}$ or $T^{\rm grav}_{r\phi}$. By the thermal balance between viscous heating and radiative cooling, we can derive $\alpha_{\rm grav}\approx 4/[9\gamma(\gamma-1)\beta]$ \citep{Armitage_2020} such that $\alpha_{\rm grav}\approx 0.1$ at $\beta\approx 1$ and $\gamma\approx 2$. This argument is for a nonlinear self-sustained disk but our analysis is about linear stability (i.e., perturbation grows but not at the equilibrium balance) so that the disk $\alpha$ in our linear study is roughly proportional to disk mass (see Eq. \eqref{eq:alpha}).

\begin{figure*}
    \centering
    \subfigure[]{
    \label{subFig:alpha_L}
    \includegraphics[width=0.3083\textwidth]{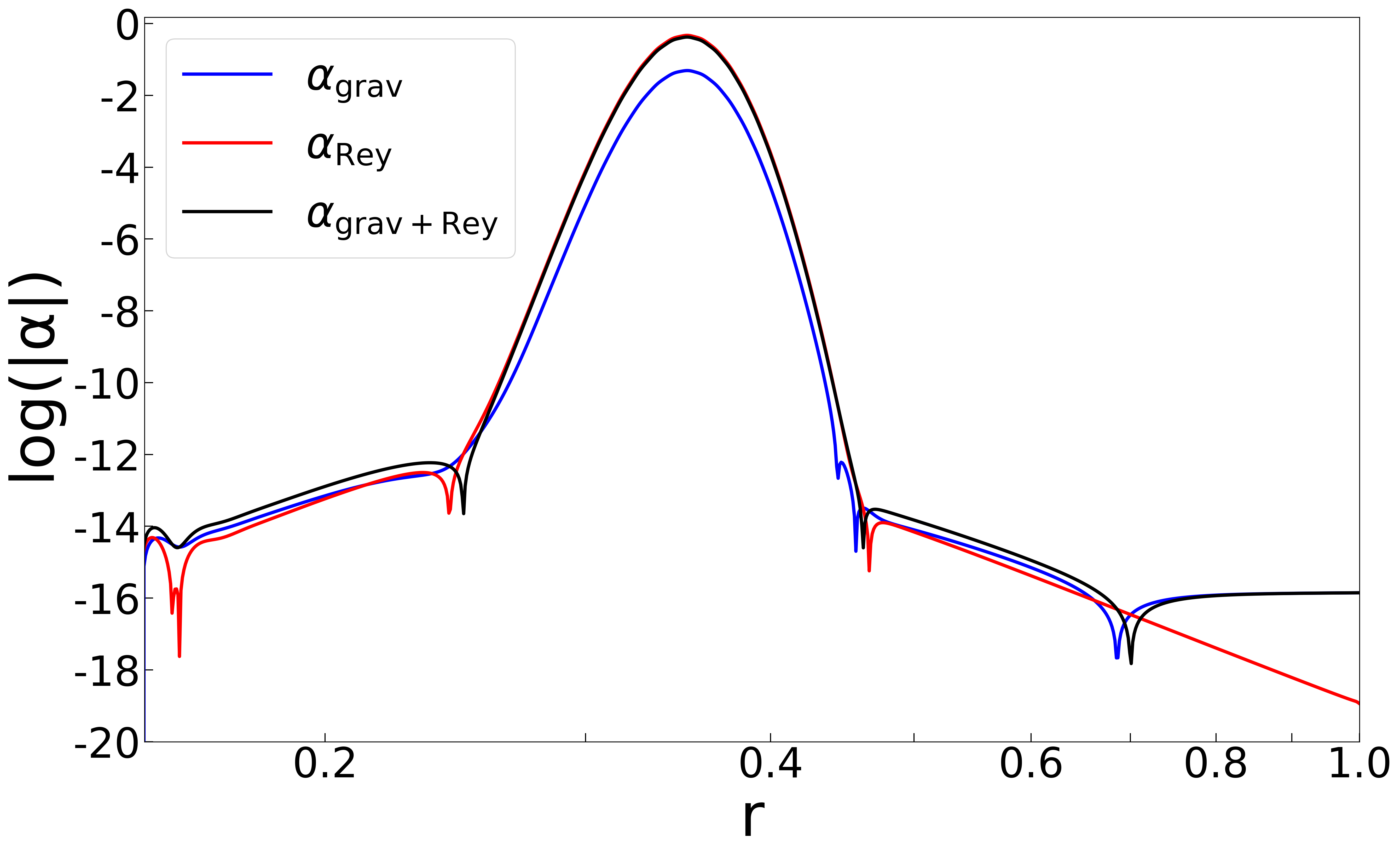}}
    \hspace{0.1cm}
    \subfigure[]{
    \label{subFig:alpha_I}
    \includegraphics[width=0.3\textwidth]{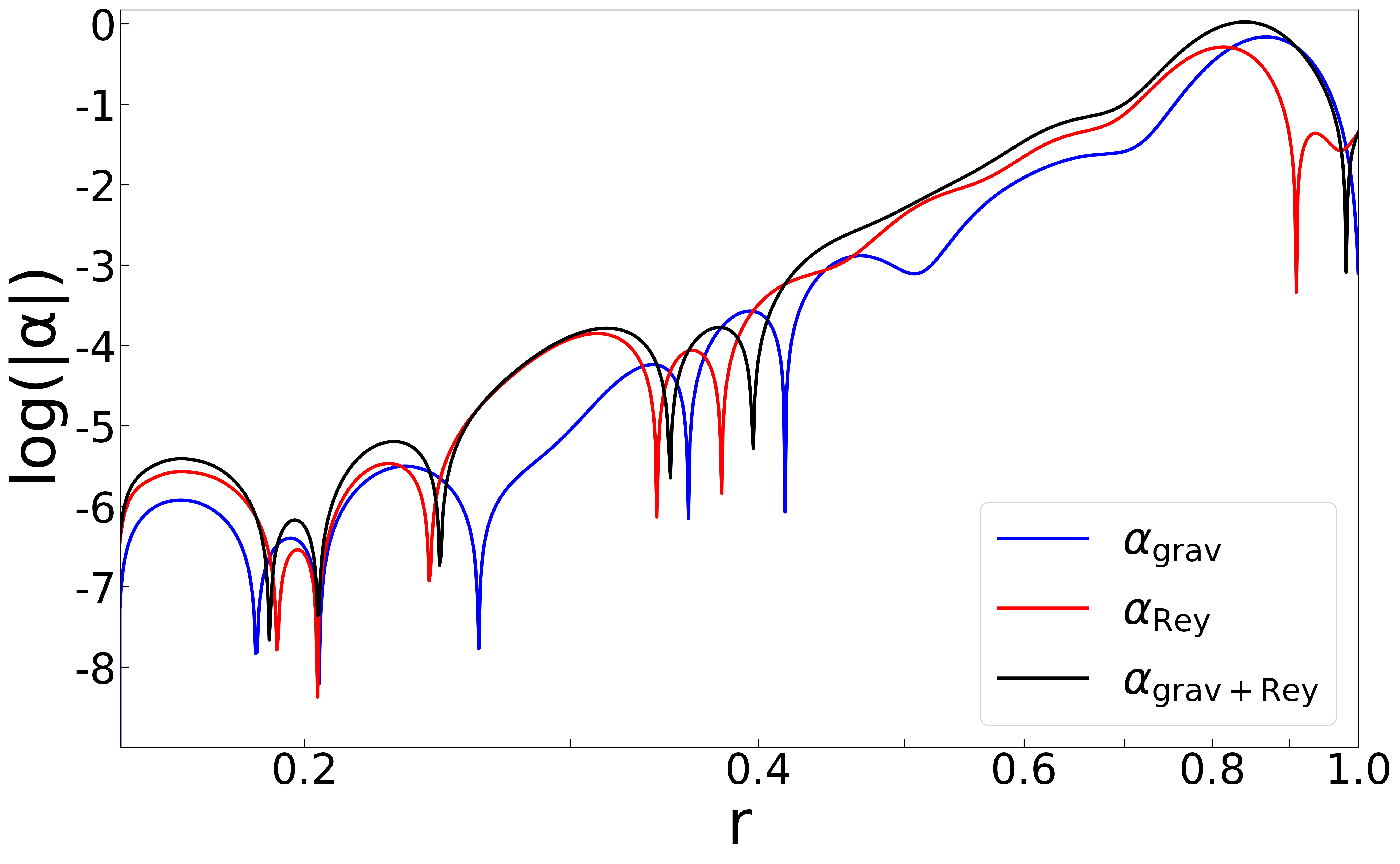}}
    \hspace{0.1cm}
    \subfigure[]{
    \label{subFig:alpha_H}
    \includegraphics[width=0.3\textwidth]{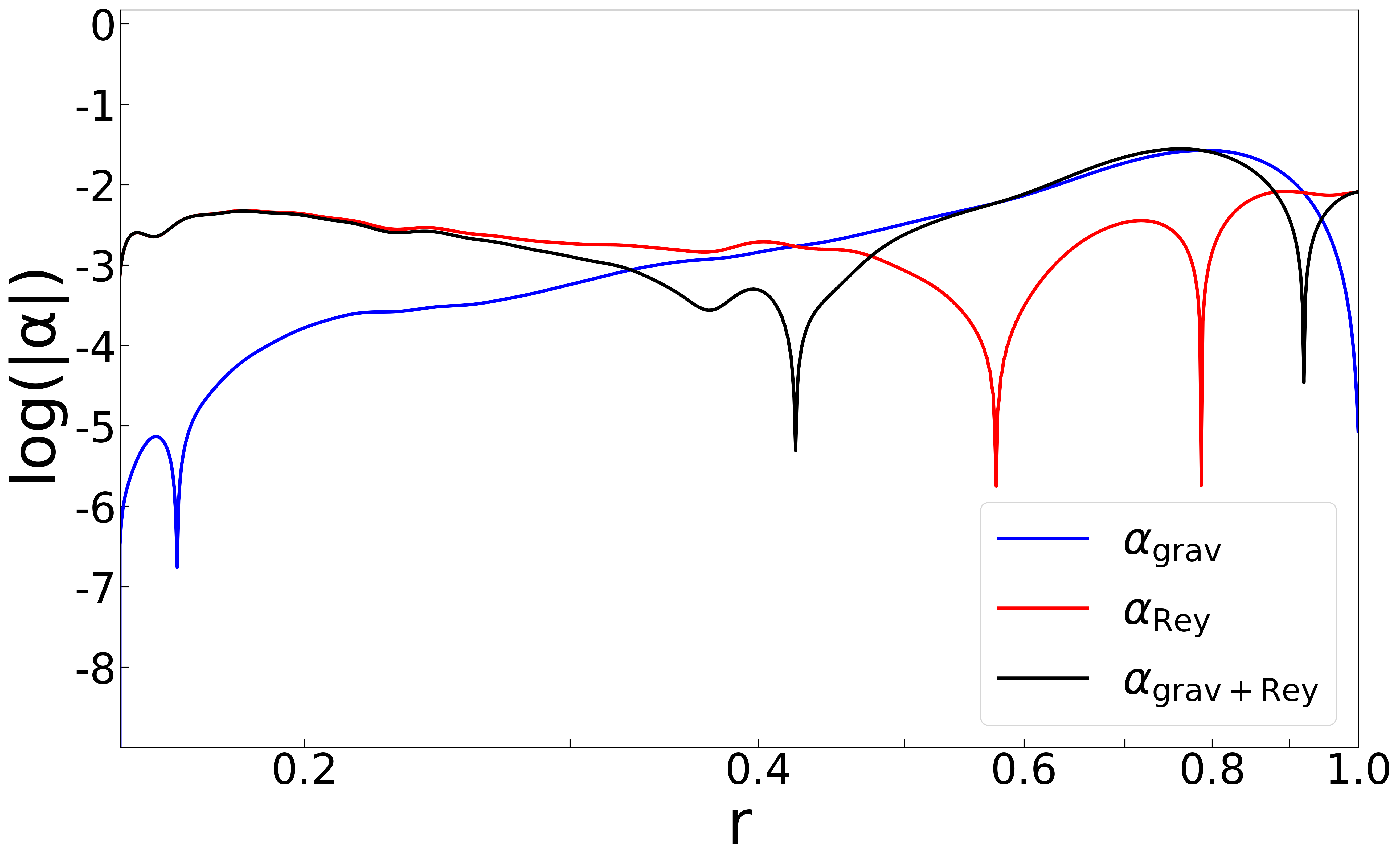}}
    \caption{Radial distribution of $\alpha_{\rm{Rey}}$, $\alpha_{\rm{grav}}$, and $\alpha_{\rm{grav+Rey}}$ in the local regime at ($\overline{Q}=0.53, \beta=1$) (a),  in the transitional regime at ($\overline Q=1.15,\beta=1$) (b), and in the global regime at $\overline{Q}=1.65, \beta=10^{-2}$ (c). The x-axis is plotted on a logarithmic scale. Red and blue solid lines represent $\alpha_{\rm{Rey}}$ and $\alpha_{\rm{grav}}$, respectively, while black solid lines show the sum.}
    \label{Fig:alpha}
\end{figure*}

Fig.\ref{Fig:alpha} shows the distribution of $\alpha_{\rm grav}$ and $\alpha_{\rm Rey}$ at different $\overline Q$ and $\beta$ in the three regimes. The deep valleys are caused by the sign reversals of torque. In our 2D study, $\alpha_{\rm grav}$ and $\alpha_{\rm Rey}$ are comparable in most regions of disk, but it is reported that $\alpha_{\rm Rey}$ is less significant than $\alpha_{\rm grav}$ in 3D studies \cite[e.g.,][]{Michael2012ApJ...746...98M,Steiman-Cameron2013ApJ...768..192S,Bae2016ApJ...829...13B,Bethune2022A&A...663A.138B}. In the local and global regimes $\alpha_{\rm grav}$ at $r_{|\delta\Sigma|_{\rm max}}$ reaches $\sim 10^{-2}$, but in the transitional regime it reaches $\sim 1$. Although density perturbation becomes highest at $r_{|\delta\Sigma|_{\rm max}}$ ($|\delta\Sigma|/\Sigma\gtrsim 50\%$ in the local regime, 20-40\% in the transitional regime and 15\% in the global regime), because of the other quantities in Eq.\eqref{eq:alpha} (e.g., $\delta\Psi$, $r$, etc.), $\alpha_{\rm grav}$ at $r_{|\delta\Sigma|_{\rm max}}$ in the transitional regime is much stronger than in the other two regimes.

Our linear studies in both local and global regimes show strong similarities with 3D simulations of disks governed by GI \citep[e.g.,][]{Boss2017ApJ...836...53B}. The rapid growth of instability in the local regime with low $\overline{Q}$ tends to create fragmentations and gravitationally bound objects in the inner disk, such as high-mass planets and brown dwarfs, regardless of cooling timescales. However, since GI saturates in the subsequent nonlinear phase and cooling is inefficient in the inner disk (e.g., $<1$ AU), fragmentation can be suppressed. For disks in the global regime, the lower growth rate leads to long-lived non-axisymmetric spirals. Notably, even for high $\overline{Q}$ ($\gtrsim 1.5$) and high $\beta$ ($\gtrsim 10$), the effective $\alpha$ can still reach $\sim 10^{-3}$, consistent with results from marginally gravitationally unstable disk studies \citep{Boss2012AREPS..40...23B}, which suggest that GI could be the driving mechanism for generic accretion disk evolution.

The transitional regime is particularly intriguing, as robust instability develops in the outer disk, where $\beta \lesssim 1$ is typically satisfied \citep{Lin2015ApJ...811...17L,Pfeil2019ApJ...871..150P}. With the efficient cooling, substructures, fragmentations and substellar companions can be created by GI. Since the transitional regime dominates at $\beta\approx 1$ (see Fig.\ref{contrast regime}), we expect that spirals with pitch angles ranging from a few degrees to over 10 degrees can be observed in the outer disk. Moreover, the large effective $\alpha_{\rm{grav}}$ in the transitional regime implies that GI is responsible for the angular momentum transport in the outer disk rather than viscosity.


\section{Summary and discussion} \label{Summary}

In this paper we explore the characteristics of GI in a protoplanetary disk at various $\overline{Q}$ and $\beta$, and our key results are summarized as follows. Firstly, faster/slower cooling timescale (smaller/larger $\beta$) leads to faster/slower growth rate, which is consistent with numerical results. Moreover, growth rate changes sharply at $\beta\approx 1$. Secondly, different $\overline Q$ corresponds to different modes: low $\overline Q$ to local modes, intermediate $\overline Q\approx 1$ to transitional modes, and high $\overline Q$ to global modes. Density perturbation maximum $r_{|{\delta\Sigma}|_{\rm max}}$ of the local and global modes is located in the inner disk and close to equilibrium density maximum $r_{\Sigma_{\rm max}}$, but $r_{|{\delta\Sigma}|_{\rm max}}$ of the transitional modes in the outer disk and far away from $r_{\Sigma_{\rm max}}$. Moreover, the transitional modes lead to much stronger $\alpha_{\rm grav}$ than the other two modes, suggesting much more efficient transport of angular momentum. Thirdly, the parameter scan reveals that the transitional modes dominate at $\beta\approx 1$. In summary, we infer that GI works in the outer disk at $\overline Q\approx 1$ and $\beta\approx 1$ to trigger the transitional modes for efficient transport of angular momentum, disk substructures (e.g., spirals and fragmentations), and planet/brown dwarf formation.

We assume a 2D thin disk. Although our calculations involve a relatively large aspect ratio, it should be noted that an early-age disk can be fairly thick even with an envelope, which will induce the additional mechanisms \citep{Kratter2016ARA&A..54..271K}. Moreover, our linear analysis cannot capture the nonlinear effects such as shocks, wave amplitude saturation, etc. In addition, some other factors such as accretion and magnetic fields can be taken into account.
\\
\\
We thank Hongping Deng and Cong Yu for the help on equilibrium, Qiang Hou for his initial assistance, Haiyang Wang for self-gravity, Yuru Xu for boundary conditions, and Cathie Clarke for disk $\alpha$.

\appendix
\section{Eigenvalue problem} \label{App1}

The basic linear perturbation equations (\ref{eqlin1})-(\ref{eqlin5}) are translated to an eigenvalue problem
\begin{equation} 
    im\Omega\delta\Sigma - \frac{1}{r}\frac{d(r\Sigma)}{dr}\delta u_{r} - \Sigma\frac{d}{dr}\delta u_{r} + \frac{im\Sigma}{r}\delta u_{\phi} = im\omega_{p}\delta\Sigma, 
\end{equation}
   
\begin{equation}
   \frac{1}{\Sigma^{2}}\frac{dP}{dr}\delta\Sigma - \frac{1}{\Sigma}\frac{d}{dr}\delta P + im\Omega\delta u_{r} + 2\Omega\delta u_{\phi} - \frac{d}{dr}\Phi_{d} = im\omega_{p}\delta u_{r}, 
\end{equation}
   
\begin{equation}
   \frac{im}{r\Sigma}\delta P - \frac{\kappa^{2}}{2\Omega}\delta u_{r} + im\Omega\delta u_{\phi} + \frac{im}{r}\Phi_{d} = im\omega_{p}\delta u_{\phi}, 
\end{equation}
   
\begin{equation} 
    \frac{c_{s,adi}^{2}}{\gamma t_{c}}\delta\Sigma + \left(-\frac{1}{t_{c}}+im\Omega\right)\delta P - \left[\frac{\Sigma c_{s,adi}^{2}}{L_{s}} + \frac{c_{s,adi}^{2}}{r}\frac{d(r\Sigma)}{dr}\right]\delta u_{r} - c_{s,adi}^{2}\Sigma\frac{d}{dr}\delta u_{r} + \frac{im\Sigma c_{s,adi}^{2}}{r}\delta u_{\phi} = im\omega_{p}\delta P. 
\end{equation}
Furthermore, we write them as
\begin{equation} \label{matrix1}
    \boldsymbol{M}
    \left[\begin{matrix}
    \delta\Sigma & \delta P & \delta u_{r} & \delta u_{\phi}
    \end{matrix}\right]^T
     = \boldsymbol{i \omega_{m}}
    \left[\begin{matrix}
    \delta\Sigma & \delta P & \delta u_{r} & \delta u_{\phi}
    \end{matrix} \right]^T,
\end{equation}
\begin{equation} \label{matrix2}
    \boldsymbol{M} = \boldsymbol{M_{1}} + \boldsymbol{M_{2}} = 
    \left[\begin{matrix}
    \boldsymbol{M}_{11} & \boldsymbol{M}_{12} & \boldsymbol{M}_{13} & \boldsymbol{M}_{14} \\
    \boldsymbol{M}_{21} & \boldsymbol{M}_{22} & \boldsymbol{M}_{23} & \boldsymbol{M}_{24} \\
    \boldsymbol{M}_{31} & \boldsymbol{M}_{32} & \boldsymbol{M}_{33} & \boldsymbol{M}_{34} \\
    \boldsymbol{M}_{41} & \boldsymbol{M}_{42} & \boldsymbol{M}_{43} & \boldsymbol{M}_{44} 
    \end{matrix}\right]. 
\end{equation}
We divide the disk into $N$ grids and the matrix $\boldsymbol{M}$ will be a $4N \times 4N$ coefficient matrix. The matrix $\boldsymbol{M_{1}}$ is sparse, containing coefficients for $\delta\Sigma, \delta P, \delta u_{r}, \delta u_{\phi}$ and their first-order derivatives. For the first-order differential operators, we employ a central finite difference scheme on logarithmically spaced radial grid points
\begin{equation} \label{central difference}
    \left.\frac{dX}{dr}\right\vert_{i}=\frac{1}{r_{i}}\sum_{j} D_{ij}^{(1)} X_{j}
\end{equation}
where $D_{ij}$ represents the first-order differentiation matrices (e.g., \cite{Adams1989}). The matrix $\boldsymbol{M_{2}}$ implements $\delta\Psi_{m}$ and $d\delta\Psi_{m}/dr$ terms. We calculate $\boldsymbol{M_{2}}$ with the method used in \cite{Laughlin1996} and \cite{Lee2019} to avoid the singularity of self-gravity potential integration. Boundary conditions \eqref{IBC} and \eqref{OBC} are imposed on some rows of $\boldsymbol{M}$. It is noteworthy that we obtain $4N$ eigenvalues and eigenfunctions. However, we focus on the most unstable mode, which owns the largest growth rate.

\bibliography{paper}{}

\begin{thebibliography}{}
\expandafter\ifx\csname natexlab\endcsname\relax\def\natexlab#1{#1}\fi
\providecommand{\url}[1]{\href{#1}{#1}}
\providecommand{\dodoi}[1]{doi:~\href{http://doi.org/#1}{\nolinkurl{#1}}}
\providecommand{\doeprint}[1]{\href{http://ascl.net/#1}{\nolinkurl{http://ascl.net/#1}}}
\providecommand{\doarXiv}[1]{\href{https://arxiv.org/abs/#1}{\nolinkurl{https://arxiv.org/abs/#1}}}

\bibitem[{{Adams} \& {Lin}(1993)}]{AdamsLin1993prpl.conf..721A}
{Adams}, F.~C., \& {Lin}, D.~N.~C. 1993, in Protostars and Planets III, ed.
  E.~H. {Levy} \& J.~I. {Lunine}, 721

\bibitem[{{Adams} {et~al.}(1989){Adams}, {Ruden}, \& {Shu}}]{Adams1989}
{Adams}, F.~C., {Ruden}, S.~P., \& {Shu}, F.~H. 1989, \apj, 347, 959,
  \dodoi{10.1086/168187}

\bibitem[{{Andrews} {et~al.}(2018){Andrews}, {Huang}, {P{\'e}rez}, {Isella},
  {Dullemond}, {Kurtovic}, {Guzm{\'a}n}, {Carpenter}, {Wilner}, {Zhang}, {Zhu},
  {Birnstiel}, {Bai}, {Benisty}, {Hughes}, {{\"O}berg}, \&
  {Ricci}}]{Andrews2018}
{Andrews}, S.~M., {Huang}, J., {P{\'e}rez}, L.~M., {et~al.} 2018, \apjl, 869,
  L41, \dodoi{10.3847/2041-8213/aaf741}

\bibitem[{Armitage(2020)}]{Armitage_2020}
Armitage, P.~J. 2020, Astrophysics of Planet Formation, 2nd edn. (Cambridge
  University Press)

\bibitem[{{Avenhaus} {et~al.}(2017){Avenhaus}, {Quanz}, {Schmid}, {Dominik},
  {Stolker}, {Ginski}, {de Boer}, {Szul{\'a}gyi}, {Garufi}, {Zurlo},
  {Hagelberg}, {Benisty}, {Henning}, {M{\'e}nard}, {Meyer}, {Baruffolo},
  {Bazzon}, {Beuzit}, {Costille}, {Dohlen}, {Girard}, {Gisler}, {Kasper},
  {Mouillet}, {Pragt}, {Roelfsema}, {Salasnich}, \&
  {Sauvage}}]{Avenhaus2017AJ....154...33A}
{Avenhaus}, H., {Quanz}, S.~P., {Schmid}, H.~M., {et~al.} 2017, \aj, 154, 33,
  \dodoi{10.3847/1538-3881/aa7560}

\bibitem[{{Backus} \& {Quinn}(2016)}]{Backus2016MNRAS.463.2480B}
{Backus}, I., \& {Quinn}, T. 2016, \mnras, 463, 2480,
  \dodoi{10.1093/mnras/stw1825}

\bibitem[{{Bae} {et~al.}(2023){Bae}, {Isella}, {Zhu}, {Martin}, {Okuzumi}, \&
  {Suriano}}]{Bae2023ASPC..534..423B}
{Bae}, J., {Isella}, A., {Zhu}, Z., {et~al.} 2023, in Astronomical Society of
  the Pacific Conference Series, Vol. 534, Protostars and Planets VII, ed.
  S.~{Inutsuka}, Y.~{Aikawa}, T.~{Muto}, K.~{Tomida}, \& M.~{Tamura}, 423,
  \dodoi{10.48550/arXiv.2210.13314}

\bibitem[{{Bae} {et~al.}(2016){Bae}, {Nelson}, {Hartmann}, \&
  {Richard}}]{Bae2016ApJ...829...13B}
{Bae}, J., {Nelson}, R.~P., {Hartmann}, L., \& {Richard}, S. 2016, \apj, 829,
  13, \dodoi{10.3847/0004-637X/829/1/13}

\bibitem[{{Baehr} {et~al.}(2017){Baehr}, {Klahr}, \&
  {Kratter}}]{Baehr2017ApJ...848...40B}
{Baehr}, H., {Klahr}, H., \& {Kratter}, K.~M. 2017, \apj, 848, 40,
  \dodoi{10.3847/1538-4357/aa8a66}

\bibitem[{{Benisty} {et~al.}(2015){Benisty}, {Juhasz}, {Boccaletti},
  {Avenhaus}, {Milli}, {Thalmann}, {Dominik}, {Pinilla}, {Buenzli}, {Pohl},
  {Beuzit}, {Birnstiel}, {de Boer}, {Bonnefoy}, {Chauvin}, {Christiaens},
  {Garufi}, {Grady}, {Henning}, {Huelamo}, {Isella}, {Langlois}, {M{\'e}nard},
  {Mouillet}, {Olofsson}, {Pantin}, {Pinte}, \&
  {Pueyo}}]{Benisty2015A&A...578L...6B}
{Benisty}, M., {Juhasz}, A., {Boccaletti}, A., {et~al.} 2015, \aap, 578, L6,
  \dodoi{10.1051/0004-6361/201526011}

\bibitem[{{Benisty} {et~al.}(2017){Benisty}, {Stolker}, {Pohl}, {de Boer},
  {Lesur}, {Dominik}, {Dullemond}, {Langlois}, {Min}, {Wagner}, {Henning},
  {Juhasz}, {Pinilla}, {Facchini}, {Apai}, {van Boekel}, {Garufi}, {Ginski},
  {M{\'e}nard}, {Pinte}, {Quanz}, {Zurlo}, {Boccaletti}, {Bonnefoy}, {Beuzit},
  {Chauvin}, {Cudel}, {Desidera}, {Feldt}, {Fontanive}, {Gratton}, {Kasper},
  {Lagrange}, {LeCoroller}, {Mouillet}, {Mesa}, {Sissa}, {Vigan}, {Antichi},
  {Buey}, {Fusco}, {Gisler}, {Llored}, {Magnard}, {Moeller-Nilsson}, {Pragt},
  {Roelfsema}, {Sauvage}, \& {Wildi}}]{Benisty2017A&A...597A..42B}
{Benisty}, M., {Stolker}, T., {Pohl}, A., {et~al.} 2017, \aap, 597, A42,
  \dodoi{10.1051/0004-6361/201629798}

\bibitem[{{B{\'e}thune} \& {Latter}(2022)}]{Bethune2022A&A...663A.138B}
{B{\'e}thune}, W., \& {Latter}, H. 2022, \aap, 663, A138,
  \dodoi{10.1051/0004-6361/202243219}

\bibitem[{Binney \& Tremaine(2008)}]{Binney2008}
Binney, J., \& Tremaine, S. 2008, Galactic Dynamics: Second Edition, rev -
  revised, 2 edn. (Princeton University Press).
\newblock \url{http://www.jstor.org/stable/j.ctvc778ff}

\bibitem[{Boley \& Durisen(2010)}]{Boley_2010}
Boley, A.~C., \& Durisen, R.~H. 2010, The Astrophysical Journal, 724, 618,
  \dodoi{10.1088/0004-637X/724/1/618}

\bibitem[{{Boley} {et~al.}(2006){Boley}, {Mej{\'\i}a}, {Durisen}, {Cai},
  {Pickett}, \& {D'Alessio}}]{Boley2006ApJ...651..517B}
{Boley}, A.~C., {Mej{\'\i}a}, A.~C., {Durisen}, R.~H., {et~al.} 2006, \apj,
  651, 517, \dodoi{10.1086/507478}

\bibitem[{Boss(1997)}]{Boss1997}
Boss, A.~P. 1997, Science, 276, 1836, \dodoi{10.1126/science.276.5320.1836}

\bibitem[{{Boss}(2001)}]{Boss2001ApJ...563..367B}
{Boss}, A.~P. 2001, \apj, 563, 367, \dodoi{10.1086/323694}

\bibitem[{{Boss}(2012)}]{Boss2012AREPS..40...23B}
---. 2012, Annual Review of Earth and Planetary Sciences, 40, 23,
  \dodoi{10.1146/annurev-earth-042711-105552}

\bibitem[{{Boss}(2017)}]{Boss2017ApJ...836...53B}
---. 2017, \apj, 836, 53, \dodoi{10.3847/1538-4357/836/1/53}

\bibitem[{{Cameron}(1978)}]{Cameron1978M&P....18....5C}
{Cameron}, A.~G.~W. 1978, Moon and Planets, 18, 5, \dodoi{10.1007/BF00896696}

\bibitem[{{Canovas} {et~al.}(2018){Canovas}, {Montesinos}, {Schreiber},
  {Cieza}, {Eiroa}, {Meeus}, {de Boer}, {M{\'e}nard}, {Wahhaj},
  {Riviere-Marichalar}, {Olofsson}, {Garufi}, {Rebollido}, {van Holstein},
  {Caceres}, {Hardy}, \& {Villaver}}]{Canovas2018A&A...610A..13C}
{Canovas}, H., {Montesinos}, B., {Schreiber}, M.~R., {et~al.} 2018, \aap, 610,
  A13, \dodoi{10.1051/0004-6361/201731640}

\bibitem[{{Chen} {et~al.}(2021){Chen}, {Yu}, \& {Ho}}]{Chen2021}
{Chen}, E., {Yu}, S.-Y., \& {Ho}, L.~C. 2021, \apj, 906, 19,
  \dodoi{10.3847/1538-4357/abc7c5}

\bibitem[{{Coradini} {et~al.}(1981){Coradini}, {Magni}, \&
  {Federico}}]{Coradini1981A&A....98..173C}
{Coradini}, A., {Magni}, G., \& {Federico}, C. 1981, \aap, 98, 173

\bibitem[{{Currie} {et~al.}(2023){Currie}, {Biller}, {Lagrange}, {Marois},
  {Guyon}, {Nielsen}, {Bonnefoy}, \& {De Rosa}}]{Currie2023ASPC..534..799C}
{Currie}, T., {Biller}, B., {Lagrange}, A., {et~al.} 2023, in Astronomical
  Society of the Pacific Conference Series, Vol. 534, Protostars and Planets
  VII, ed. S.~{Inutsuka}, Y.~{Aikawa}, T.~{Muto}, K.~{Tomida}, \& M.~{Tamura},
  799, \dodoi{10.48550/arXiv.2205.05696}

\bibitem[{{Dong} {et~al.}(2015){Dong}, {Hall}, {Rice}, \& {Chiang}}]{Dong2015}
{Dong}, R., {Hall}, C., {Rice}, K., \& {Chiang}, E. 2015, \apjl, 812, L32,
  \dodoi{10.1088/2041-8205/812/2/L32}

\bibitem[{{Durisen} {et~al.}(2007){Durisen}, {Boss}, {Mayer}, {Nelson},
  {Quinn}, \& {Rice}}]{Durisen2007prpl.conf..607D}
{Durisen}, R.~H., {Boss}, A.~P., {Mayer}, L., {et~al.} 2007, in Protostars and
  Planets V, ed. B.~{Reipurth}, D.~{Jewitt}, \& K.~{Keil}, 607,
  \dodoi{10.48550/arXiv.astro-ph/0603179}

\bibitem[{{Forgan} {et~al.}(2015){Forgan}, {Parker}, \&
  {Rice}}]{Forgan2015MNRAS.447..836F}
{Forgan}, D., {Parker}, R.~J., \& {Rice}, K. 2015, \mnras, 447, 836,
  \dodoi{10.1093/mnras/stu2504}

\bibitem[{{Forgan} \& {Rice}(2013)}]{Forgan2013MNRAS.432.3168F}
{Forgan}, D., \& {Rice}, K. 2013, \mnras, 432, 3168,
  \dodoi{10.1093/mnras/stt672}

\bibitem[{{Gammie}(2001)}]{Gammie2001}
{Gammie}, C.~F. 2001, \apj, 553, 174, \dodoi{10.1086/320631}

\bibitem[{{Garufi} {et~al.}(2013){Garufi}, {Quanz}, {Avenhaus}, {Buenzli},
  {Dominik}, {Meru}, {Meyer}, {Pinilla}, {Schmid}, \&
  {Wolf}}]{Garufi2013A&A...560A.105G}
{Garufi}, A., {Quanz}, S.~P., {Avenhaus}, H., {et~al.} 2013, \aap, 560, A105,
  \dodoi{10.1051/0004-6361/201322429}

\bibitem[{{Goldreich} \& {Lynden-Bell}(1965)}]{Goldreich1965}
{Goldreich}, P., \& {Lynden-Bell}, D. 1965, \mnras, 130, 125,
  \dodoi{10.1093/mnras/130.2.125}

\bibitem[{{Goodman} \& {Narayan}(1988)}]{Goodman1988}
{Goodman}, J., \& {Narayan}, R. 1988, \mnras, 231, 97,
  \dodoi{10.1093/mnras/231.1.97}

\bibitem[{{Hadley} {et~al.}(2014){Hadley}, {Fernandez}, {Imamura}, {Keever},
  {Tumblin}, \& {Dumas}}]{Hadley2014Ap&SS.353..191H}
{Hadley}, K.~Z., {Fernandez}, P., {Imamura}, J.~N., {et~al.} 2014, \apss, 353,
  191, \dodoi{10.1007/s10509-014-1994-8}

\bibitem[{{Hall} {et~al.}(2019){Hall}, {Dong}, {Rice}, {Harries}, {Najita},
  {Alexander}, \& {Brittain}}]{Hall2019ApJ...871..228H}
{Hall}, C., {Dong}, R., {Rice}, K., {et~al.} 2019, \apj, 871, 228,
  \dodoi{10.3847/1538-4357/aafac2}

\bibitem[{{Hall} {et~al.}(2018){Hall}, {Rice}, {Dipierro}, {Forgan}, {Harries},
  \& {Alexander}}]{Hall2018MNRAS.477.1004H}
{Hall}, C., {Rice}, K., {Dipierro}, G., {et~al.} 2018, \mnras, 477, 1004,
  \dodoi{10.1093/mnras/sty550}

\bibitem[{{Huang} {et~al.}(2018){Huang}, {Andrews}, {P{\'e}rez}, {Zhu},
  {Dullemond}, {Isella}, {Benisty}, {Bai}, {Birnstiel}, {Carpenter},
  {Guzm{\'a}n}, {Hughes}, {{\"O}berg}, {Ricci}, {Wilner}, \&
  {Zhang}}]{Huang2018ApJ...869L..43H}
{Huang}, J., {Andrews}, S.~M., {P{\'e}rez}, L.~M., {et~al.} 2018, \apjl, 869,
  L43, \dodoi{10.3847/2041-8213/aaf7a0}

\bibitem[{{Jeans}(1902)}]{Jeans1902}
{Jeans}, J.~H. 1902, Philosophical Transactions of the Royal Society of London
  Series A, 199, 1, \dodoi{10.1098/rsta.1902.0012}

\bibitem[{{Johnson} \& {Gammie}(2003)}]{Johnson2003ApJ...597..131J}
{Johnson}, B.~M., \& {Gammie}, C.~F. 2003, \apj, 597, 131,
  \dodoi{10.1086/378392}

\bibitem[{{Kratter} \& {Lodato}(2016)}]{Kratter2016ARA&A..54..271K}
{Kratter}, K., \& {Lodato}, G. 2016, \araa, 54, 271,
  \dodoi{10.1146/annurev-astro-081915-023307}

\bibitem[{{Kratter} {et~al.}(2010){Kratter}, {Murray-Clay}, \&
  {Youdin}}]{Kratter2010ApJ...710.1375K}
{Kratter}, K.~M., {Murray-Clay}, R.~A., \& {Youdin}, A.~N. 2010, \apj, 710,
  1375, \dodoi{10.1088/0004-637X/710/2/1375}

\bibitem[{{Kurtovic} {et~al.}(2018){Kurtovic}, {P{\'e}rez}, {Benisty}, {Zhu},
  {Zhang}, {Huang}, {Andrews}, {Dullemond}, {Isella}, {Bai}, {Carpenter},
  {Guzm{\'a}n}, {Ricci}, \& {Wilner}}]{Kurtovic2018ApJ...869L..44K}
{Kurtovic}, N.~T., {P{\'e}rez}, L.~M., {Benisty}, M., {et~al.} 2018, \apjl,
  869, L44, \dodoi{10.3847/2041-8213/aaf746}

\bibitem[{{Laughlin} \& {Rozyczka}(1996)}]{Laughlin1996}
{Laughlin}, G., \& {Rozyczka}, M. 1996, \apj, 456, 279, \dodoi{10.1086/176648}

\bibitem[{{Lee} {et~al.}(2019){Lee}, {Dempsey}, \& {Lithwick}}]{Lee2019}
{Lee}, W.-K., {Dempsey}, A.~M., \& {Lithwick}, Y. 2019, \apj, 872, 184,
  \dodoi{10.3847/1538-4357/ab010c}

\bibitem[{{Lin} \& {Shu}(1964)}]{Lin1964}
{Lin}, C.~C., \& {Shu}, F.~H. 1964, \apj, 140, 646, \dodoi{10.1086/147955}

\bibitem[{{Lin} \& {Pringle}(1987)}]{Lin1987MNRAS.225..607L}
{Lin}, D.~N.~C., \& {Pringle}, J.~E. 1987, \mnras, 225, 607,
  \dodoi{10.1093/mnras/225.3.607}

\bibitem[{{Lin} \& {Youdin}(2015)}]{Lin2015ApJ...811...17L}
{Lin}, M.-K., \& {Youdin}, A.~N. 2015, \apj, 811, 17,
  \dodoi{10.1088/0004-637X/811/1/17}

\bibitem[{{Lodato} \& {Rice}(2004)}]{Lodato2004MNRAS.351..630L}
{Lodato}, G., \& {Rice}, W.~K.~M. 2004, \mnras, 351, 630,
  \dodoi{10.1111/j.1365-2966.2004.07811.x}

\bibitem[{Lodato \& Rice(2005)}]{Lodato201510.1111/j.1365-2966.2005.08875.x}
Lodato, G., \& Rice, W. K.~M. 2005, Monthly Notices of the Royal Astronomical
  Society, 358, 1489, \dodoi{10.1111/j.1365-2966.2005.08875.x}

\bibitem[{{Longarini} {et~al.}(2023){Longarini}, {Lodato}, {Bertin}, \&
  {Armitage}}]{Longarini2023MNRAS.519.2017L}
{Longarini}, C., {Lodato}, G., {Bertin}, G., \& {Armitage}, P.~J. 2023, \mnras,
  519, 2017, \dodoi{10.1093/mnras/stac3653}

\bibitem[{{Meru} {et~al.}(2017){Meru}, {Juh{\'a}sz}, {Ilee}, {Clarke},
  {Rosotti}, \& {Booth}}]{Meru2017ApJ...839L..24M}
{Meru}, F., {Juh{\'a}sz}, A., {Ilee}, J.~D., {et~al.} 2017, \apjl, 839, L24,
  \dodoi{10.3847/2041-8213/aa6837}

\bibitem[{{Michael} {et~al.}(2012){Michael}, {Steiman-Cameron}, {Durisen}, \&
  {Boley}}]{Michael2012ApJ...746...98M}
{Michael}, S., {Steiman-Cameron}, T.~Y., {Durisen}, R.~H., \& {Boley}, A.~C.
  2012, \apj, 746, 98, \dodoi{10.1088/0004-637X/746/1/98}

\bibitem[{{Michikoshi} {et~al.}(2012){Michikoshi}, {Kokubo}, \&
  {Inutsuka}}]{Michikoshi2012ApJ...746...35M}
{Michikoshi}, S., {Kokubo}, E., \& {Inutsuka}, S.-i. 2012, \apj, 746, 35,
  \dodoi{10.1088/0004-637X/746/1/35}

\bibitem[{{Miranda} \& {Rafikov}(2020)}]{Miranda2020}
{Miranda}, R., \& {Rafikov}, R.~R. 2020, \apj, 892, 65,
  \dodoi{10.3847/1538-4357/ab791a}

\bibitem[{Nayakshin(2017)}]{Nayakshin_2017}
Nayakshin, S. 2017, Publications of the Astronomical Society of Australia, 34,
  e002, \dodoi{10.1017/pasa.2016.55}

\bibitem[{{Nielsen} {et~al.}(2019){Nielsen}, {De Rosa}, {Macintosh}, {Wang},
  {Ruffio}, {Chiang}, {Marley}, {Saumon}, {Savransky}, {Ammons}, {Bailey},
  {Barman}, {Blain}, {Bulger}, {Burrows}, {Chilcote}, {Cotten}, {Czekala},
  {Doyon}, {Duch{\^e}ne}, {Esposito}, {Fabrycky}, {Fitzgerald}, {Follette},
  {Fortney}, {Gerard}, {Goodsell}, {Graham}, {Greenbaum}, {Hibon}, {Hinkley},
  {Hirsch}, {Hom}, {Hung}, {Dawson}, {Ingraham}, {Kalas}, {Konopacky},
  {Larkin}, {Lee}, {Lin}, {Maire}, {Marchis}, {Marois}, {Metchev},
  {Millar-Blanchaer}, {Morzinski}, {Oppenheimer}, {Palmer}, {Patience},
  {Perrin}, {Poyneer}, {Pueyo}, {Rafikov}, {Rajan}, {Rameau}, {Rantakyr{\"o}},
  {Ren}, {Schneider}, {Sivaramakrishnan}, {Song}, {Soummer}, {Tallis},
  {Thomas}, {Ward-Duong}, \& {Wolff}}]{Nielsen2019AJ....158...13N}
{Nielsen}, E.~L., {De Rosa}, R.~J., {Macintosh}, B., {et~al.} 2019, \aj, 158,
  13, \dodoi{10.3847/1538-3881/ab16e9}

\bibitem[{{Noh} {et~al.}(1991){Noh}, {Vishniac}, \&
  {Cochran}}]{Noh1991ApJ...383..372N}
{Noh}, H., {Vishniac}, E.~T., \& {Cochran}, W.~D. 1991, \apj, 383, 372,
  \dodoi{10.1086/170795}

\bibitem[{{Paardekooper}(2012)}]{Paardekooper2012MNRAS.421.3286P}
{Paardekooper}, S.-J. 2012, \mnras, 421, 3286,
  \dodoi{10.1111/j.1365-2966.2012.20553.x}

\bibitem[{{P{\'e}rez} {et~al.}(2016){P{\'e}rez}, {Carpenter}, {Andrews},
  {Ricci}, {Isella}, {Linz}, {Sargent}, {Wilner}, {Henning}, {Deller},
  {Chandler}, {Dullemond}, {Lazio}, {Menten}, {Corder}, {Storm}, {Testi},
  {Tazzari}, {Kwon}, {Calvet}, {Greaves}, {Harris}, \&
  {Mundy}}]{Perez2016Sci...353.1519P}
{P{\'e}rez}, L.~M., {Carpenter}, J.~M., {Andrews}, S.~M., {et~al.} 2016,
  Science, 353, 1519, \dodoi{10.1126/science.aaf8296}

\bibitem[{{Pfeil} \& {Klahr}(2019)}]{Pfeil2019ApJ...871..150P}
{Pfeil}, T., \& {Klahr}, H. 2019, \apj, 871, 150,
  \dodoi{10.3847/1538-4357/aaf962}

\bibitem[{{Pierens}(2021)}]{Pierens2021MNRAS.504.4522P}
{Pierens}, A. 2021, \mnras, 504, 4522, \dodoi{10.1093/mnras/stab183}

\bibitem[{Rice \& Nayakshin(2017)}]{Rice201710.1093/mnras/stx3255}
Rice, K., \& Nayakshin, S. 2017, Monthly Notices of the Royal Astronomical
  Society, 475, 921, \dodoi{10.1093/mnras/stx3255}

\bibitem[{{Rice} {et~al.}(2005){Rice}, {Lodato}, \&
  {Armitage}}]{Rice2005MNRAS.364L..56R}
{Rice}, W.~K.~M., {Lodato}, G., \& {Armitage}, P.~J. 2005, \mnras, 364, L56,
  \dodoi{10.1111/j.1745-3933.2005.00105.x}

\bibitem[{{Shakura} \& {Sunyaev}(1973)}]{Shakura1973A&A....24..337S}
{Shakura}, N.~I., \& {Sunyaev}, R.~A. 1973, \aap, 24, 337

\bibitem[{{Shariff} \& {Cuzzi}(2011)}]{Shariff2011ApJ...738...73S}
{Shariff}, K., \& {Cuzzi}, J.~N. 2011, \apj, 738, 73,
  \dodoi{10.1088/0004-637X/738/1/73}

\bibitem[{{Shu}(1992)}]{Shu1992}
{Shu}, F.~H. 1992, {The physics of astrophysics. Volume II: Gas dynamics.}
  (University Science books)

\bibitem[{{Speedie} {et~al.}(2024){Speedie}, {Dong}, {Hall}, {Longarini},
  {Veronesi}, {Paneque-Carre{\~n}o}, {Lodato}, {Tang}, {Teague}, \&
  {Hashimoto}}]{Speedie2024Natur.633...58S}
{Speedie}, J., {Dong}, R., {Hall}, C., {et~al.} 2024, \nat, 633, 58,
  \dodoi{10.1038/s41586-024-07877-0}

\bibitem[{{Steiman-Cameron} {et~al.}(2023){Steiman-Cameron}, {Durisen},
  {Boley}, {Michael}, {Desai}, \&
  {McConnell}}]{Steiman-Cameron2023ApJ...958..139S}
{Steiman-Cameron}, T.~Y., {Durisen}, R.~H., {Boley}, A.~C., {et~al.} 2023,
  \apj, 958, 139, \dodoi{10.3847/1538-4357/acff6d}

\bibitem[{{Steiman-Cameron} {et~al.}(2013){Steiman-Cameron}, {Durisen},
  {Boley}, {Michael}, \& {McConnell}}]{Steiman-Cameron2013ApJ...768..192S}
{Steiman-Cameron}, T.~Y., {Durisen}, R.~H., {Boley}, A.~C., {Michael}, S., \&
  {McConnell}, C.~R. 2013, \apj, 768, 192, \dodoi{10.1088/0004-637X/768/2/192}

\bibitem[{{Takeuchi} \& {Ida}(2012)}]{Takeuchi2012ApJ...749...89T}
{Takeuchi}, T., \& {Ida}, S. 2012, \apj, 749, 89,
  \dodoi{10.1088/0004-637X/749/1/89}

\bibitem[{{Tomida} {et~al.}(2017){Tomida}, {Machida}, {Hosokawa}, {Sakurai}, \&
  {Lin}}]{Tomida2017ApJ...835L..11T}
{Tomida}, K., {Machida}, M.~N., {Hosokawa}, T., {Sakurai}, Y., \& {Lin}, C.~H.
  2017, \apjl, 835, L11, \dodoi{10.3847/2041-8213/835/1/L11}

\bibitem[{{Tominaga} {et~al.}(2020){Tominaga}, {Takahashi}, \&
  {Inutsuka}}]{Tominaga2020ApJ...900..182T}
{Tominaga}, R.~T., {Takahashi}, S.~Z., \& {Inutsuka}, S.-i. 2020, \apj, 900,
  182, \dodoi{10.3847/1538-4357/abad36}

\bibitem[{{Toomre}(1964)}]{Toomre1964}
{Toomre}, A. 1964, \apj, 139, 1217, \dodoi{10.1086/147861}

\bibitem[{{Uyama} {et~al.}(2018){Uyama}, {Hashimoto}, {Muto}, {Akiyama},
  {Dong}, {de Leon}, {Sakon}, {Kudo}, {Kusakabe}, {Kuzuhara}, {Bonnefoy},
  {Abe}, {Brandner}, {Brandt}, {Carson}, {Currie}, {Egner}, {Feldt}, {Fung},
  {Goto}, {Grady}, {Guyon}, {Hayano}, {Hayashi}, {Hayashi}, {Henning},
  {Hodapp}, {Ishii}, {Iye}, {Janson}, {Kandori}, {Knapp}, {Kwon}, {Matsuo},
  {Mayama}, {Mcelwain}, {Miyama}, {Morino}, {Moro-Martin}, {Nishimura}, {Pyo},
  {Serabyn}, {Sitko}, {Suenaga}, {Suto}, {Suzuki}, {Takahashi}, {Takami},
  {Takato}, {Terada}, {Thalmann}, {Turner}, {Watanabe}, {Wisniewski}, {Yamada},
  {Yang}, {Takami}, {Usuda}, \& {Tamura}}]{Uyama2018AJ....156...63U}
{Uyama}, T., {Hashimoto}, J., {Muto}, T., {et~al.} 2018, \aj, 156, 63,
  \dodoi{10.3847/1538-3881/aacbd1}

\bibitem[{{Vigan} {et~al.}(2021){Vigan}, {Fontanive}, {Meyer}, {Biller},
  {Bonavita}, {Feldt}, {Desidera}, {Marleau}, {Emsenhuber}, {Galicher}, {Rice},
  {Forgan}, {Mordasini}, {Gratton}, {Le Coroller}, {Maire}, {Cantalloube},
  {Chauvin}, {Cheetham}, {Hagelberg}, {Lagrange}, {Langlois}, {Bonnefoy},
  {Beuzit}, {Boccaletti}, {D'Orazi}, {Delorme}, {Dominik}, {Henning}, {Janson},
  {Lagadec}, {Lazzoni}, {Ligi}, {Menard}, {Mesa}, {Messina}, {Moutou},
  {M{\"u}ller}, {Perrot}, {Samland}, {Schmid}, {Schmidt}, {Sissa}, {Turatto},
  {Udry}, {Zurlo}, {Abe}, {Antichi}, {Asensio-Torres}, {Baruffolo}, {Baudoz},
  {Baudrand}, {Bazzon}, {Blanchard}, {Bohn}, {Brown Sevilla}, {Carbillet},
  {Carle}, {Cascone}, {Charton}, {Claudi}, {Costille}, {De Caprio},
  {Delboulb{\'e}}, {Dohlen}, {Engler}, {Fantinel}, {Feautrier}, {Fusco},
  {Gigan}, {Girard}, {Giro}, {Gisler}, {Gluck}, {Gry}, {Hubin}, {Hugot},
  {Jaquet}, {Kasper}, {Le Mignant}, {Llored}, {Madec}, {Magnard}, {Martinez},
  {Maurel}, {M{\"o}ller-Nilsson}, {Mouillet}, {Moulin}, {Orign{\'e}}, {Pavlov},
  {Perret}, {Petit}, {Pragt}, {Puget}, {Rabou}, {Ramos}, {Rickman}, {Rigal},
  {Rochat}, {Roelfsema}, {Rousset}, {Roux}, {Salasnich}, {Sauvage}, {Sevin},
  {Soenke}, {Stadler}, {Suarez}, {Wahhaj}, {Weber}, \&
  {Wildi}}]{Vigan2021A&A...651A..72V}
{Vigan}, A., {Fontanive}, C., {Meyer}, M., {et~al.} 2021, \aap, 651, A72,
  \dodoi{10.1051/0004-6361/202038107}

\bibitem[{{Wagner} {et~al.}(2015){Wagner}, {Apai}, {Kasper}, \&
  {Robberto}}]{Wagner2015ApJ...813L...2W}
{Wagner}, K., {Apai}, D., {Kasper}, M., \& {Robberto}, M. 2015, \apjl, 813, L2,
  \dodoi{10.1088/2041-8205/813/1/L2}

\bibitem[{{Ward}(2000)}]{Ward2000orem.book...75W}
{Ward}, W.~R. 2000, in Origin of the Earth and Moon, ed. R.~M. {Canup},
  K.~{Righter}, \& {et al.} (University of Arizona Press), 75--84

\bibitem[{{Xu} \& {Kunz}(2021)}]{Xu2021}
{Xu}, W., \& {Kunz}, M.~W. 2021, \mnras, 508, 2142,
  \dodoi{10.1093/mnras/stab2715}

\bibitem[{{Youdin}(2005)}]{Youdin2005astro.ph..8659Y}
{Youdin}, A.~N. 2005, arXiv e-prints, astro,
  \dodoi{10.48550/arXiv.astro-ph/0508659}

\end{thebibliography}
\bibliographystyle{aasjournal}

\end{document}